\documentclass[journal,comsoc]{IEEEtran}

\usepackage[T1]{fontenc}%

\ifCLASSINFOpdf
  \usepackage[pdftex]{graphicx}
\else
\fi

\usepackage{amsmath}
\usepackage[cmintegrals]{newtxmath}

\newif\ifshowcomments
\showcommentstrue

\ifshowcomments
\newcommand{\mynote}[2]{\textcolor{blue}{\fbox{\bfseries\sffamily\scriptsize#1}}
  \textcolor{blue}{{$/*$\textsf{\emph{#2}}$*/$}}}
\newcommand{\mikenote}[2]{\textcolor{red}{\fbox{\bfseries\sffamily\scriptsize#1}}
  \textcolor{red}{{$/*$\textsf{\emph{#2}}$*/$}}}
\newcommand{\pynote}[2]{\textcolor{cyan}{\fbox{\bfseries\sffamily\scriptsize#1}}
  \textcolor{cyan}{{$/*$\textsf{\emph{#2}}$*/$}}}
\newcommand{\cznote}[2]{\textcolor{magenta}{\fbox{\bfseries\sffamily\scriptsize#1}}
  \textcolor{magenta}{{$/*$\textsf{\emph{#2}}$*/$}}}
\else
\newcommand{\mynote}[2]{}
\newcommand{\pynote}[2]{}
\newcommand{\mikenote}[2]{}
\newcommand{\cznote}[2]{}
\fi

\usepackage{textcomp}
\usepackage{xcolor}
\usepackage[final]{changes}

\usepackage[ruled,vlined]{algorithm2e}
\usepackage{tabularx}
\usepackage{booktabs}
\usepackage{amsmath}
\usepackage{comment}
\usepackage{xspace}
\usepackage{subcaption}
\usepackage{enumitem}

\newcommand{\AlgoName}{IRONWAN\xspace}
\newcommand{\Arr}{RMIP\xspace}

\newcommand{\IP}{InterPred\xspace}

\newtheorem{sub-problem}{Sub-Problem}

\begin{document}
\title{IRONWAN: Increasing Reliability of Overlapping Networks in LoRaWAN}
\author{Laksh Bhatia,
Po-Yu Chen,
Michael Breza,
Cong Zhao,
Julie A. McCann\\
\textit{Department of Computing, Imperial College London}\\
\textit{
\{laksh.bhatia16,po-yu.chen11,michael.breza04,c.zhao,j.mccann\}@imperial.ac.uk}
}

\markboth{Journal of \LaTeX\ Class Files,~Vol.~14, No.~8, August~2015}%
{Shell \MakeLowercase{\textit{et al.}}: Bare Demo of IEEEtran.cls for IEEE Communications Society Journals}

\maketitle

\begin{abstract}
LoRaWAN deployments follow an ad-hoc deployment model that has organically led to overlapping communication networks, sharing the wireless spectrum, and completely unaware of each other. LoRaWAN uses ALOHA-style communication where it is almost impossible to schedule transmission between networks belonging to different owners properly. The inability to schedule overlapping networks will cause inter-network interference, which will increase node-to-gateway message losses and gateway-to-node acknowledgement failures. This problem is likely to get worse as the number of LoRaWAN networks increase. 
In response to this problem, we propose IRONWAN, a wireless overlay network that shares communication resources without modifications to underlying protocols. It utilises the broadcast nature of radio communication and enables gateway-to-gateway communication to facilitate the search for failed messages and transmit failed acknowledgements already received and cached in overlapping network's gateways. IRONWAN uses two novel algorithms, a Real-time Message Inter-arrival Predictor, to highlight when a server has not received an expected uplink message. The Interference Predictor ensures that extra gateway-to-gateway communication does not negatively impact communication bandwidth. We evaluate IRONWAN on a 1000-node simulator with up to ten gateways and a 10-node testbed with 2-gateways. Results show that IRONWAN can achieve up to 12\% higher packet delivery ratio (PDR) and total messages received per node while increasing the minimum PDR by up to 28\%. These improvements save up to 50\% node's energy. Finally, we demonstrate that IRONWAN has comparable performance to an optimal solution (wired, centralised) but with 2-32 times lower communication costs. IRONWAN also has up to 14\% better PDR when compared to FLIP, a wired-distributed gateway-to-gateway protocol in certain scenarios.
\end{abstract}


\IEEEpeerreviewmaketitle

\section{Introduction}
\label{intro}

LoRaWAN \cite{LoRaWAN} is a widely deployed Low Power Wide Area Network (LPWAN)
wireless communication protocol used in many large scale Internet of Things
(IoT) systems, including city-scale sensing, smart urban infrastructure,
precision agriculture, and industry 4.0 \cite{mekki2019comparative}.  It
provides a low-power solution to applications that tolerate low data rates, are
uplink-heavy and generally delay-tolerant. LoRaWAN is a
license-free protocol that allows users to deploy their networks (including
nodes, gateways, and servers) anywhere and in any density that they require. 

This freedom of deployment by different stakeholders in a space creates communication interference among overlapping networks or networks in close
physical proximity. LPWANs like NB-IoT and Sigfox solve the problem of overlapping networks by being carrier controlled. NB-IoT gateways are positioned to implement purposefully designed minimally overlapping cells. Overlapping may exist with SigFox, but a network's location and capacity are constrained to a single carrier. 
The carrier controlled approach constrains networks to areas where, for example, a network operator can provide communication coverage. 

The deployment model of LoRaWAN is one reason for its popularity and broad adoption. 
Over 1.3 million public and private LoRaWAN gateways have been deployed by
2021\footnote{https://www.semtech.com/lora}. LoRaWAN allows private stakeholders to build personal LPWAN networks for security and privacy reasons \cite{mekki2019comparative}. LoRaWAN is also preferred in areas where network infrastructures and Internet access are unreliable or not readily available. It can be deployed relatively cheaply without the need for a carrier. 

Overlapping LoRaWAN networks cause unexpected packet collisions and duty-cycle exhaustion. We demonstrate this in Sec. \ref{sec:background} through a simulation of a 1,000 node network with 6 gateways (results in Fig.\ref{fig:partitioning}). Our simulation shows that the packet delivery ratio (PDR) of four overlapping networks is 50\% less than that of the same sized deployment owned by a single
network. 

Without a carrier-controlled system, it is a challenge for overlapping LoRaWAN deployments to optimise their network settings (e.g. scheduling, transmission parameters) as in Sigfox and NB-IoT. Simply sharing network metrics (e.g. network loads and node transmission patterns) between networks is not sufficient to solve this problem without sophisticated analysis to determine when gateways should interact and how to do so without disrupting the other networks. A naive gateway-to-gateway (G2G) solution would create potential security vulnerabilities. 
Malicious users could send falsified information for their benefit or jam the
network at critical times \cite{Aras_2017}.

Recent research has shed light on the problem of overlapping networks in
LoRaWAN and exposed beneficial opportunities \added[]{\cite{FLIP,dongare2018charm,Liu2020}}. LoRaWAN operates in the
unlicensed spectrum, and gateways receive all messages in their communication range. LoRaWAN encrypts the messages so that only the desired users can decode them. These encrypted messages, however, are available at gateways and servers that have no use for them. This presents an opportunity for gateways and servers to deliver messages
not destined for them to other networks that may have missed those messages. 

For example, \cite{packetbroker} proposes message exchanging between network servers via an Internet-based cloud service. FLIP \cite{FLIP} constructs a G2G backhaul network to transfer the reception and acknowledgement responsibility of nodes from one network to gateways of an overlapping network. However, these approaches require reliable wired backhaul access (for internet access or G2G communications), which may not be available in hazardous or remote areas
\cite{unreliableinternet}. 
Furthermore, these approaches (e.g.\cite{FLIP}) require the network owners to co-ordinate and agree to exchange authentication keys or grant full access \cite{Dwijaksara2019} to the gateways of other networks. This puts each network at risk of injection attacks or falsified data and the dropping of uplink or downlink messages.

\begin{figure}
  \includegraphics[width=0.5\textwidth]{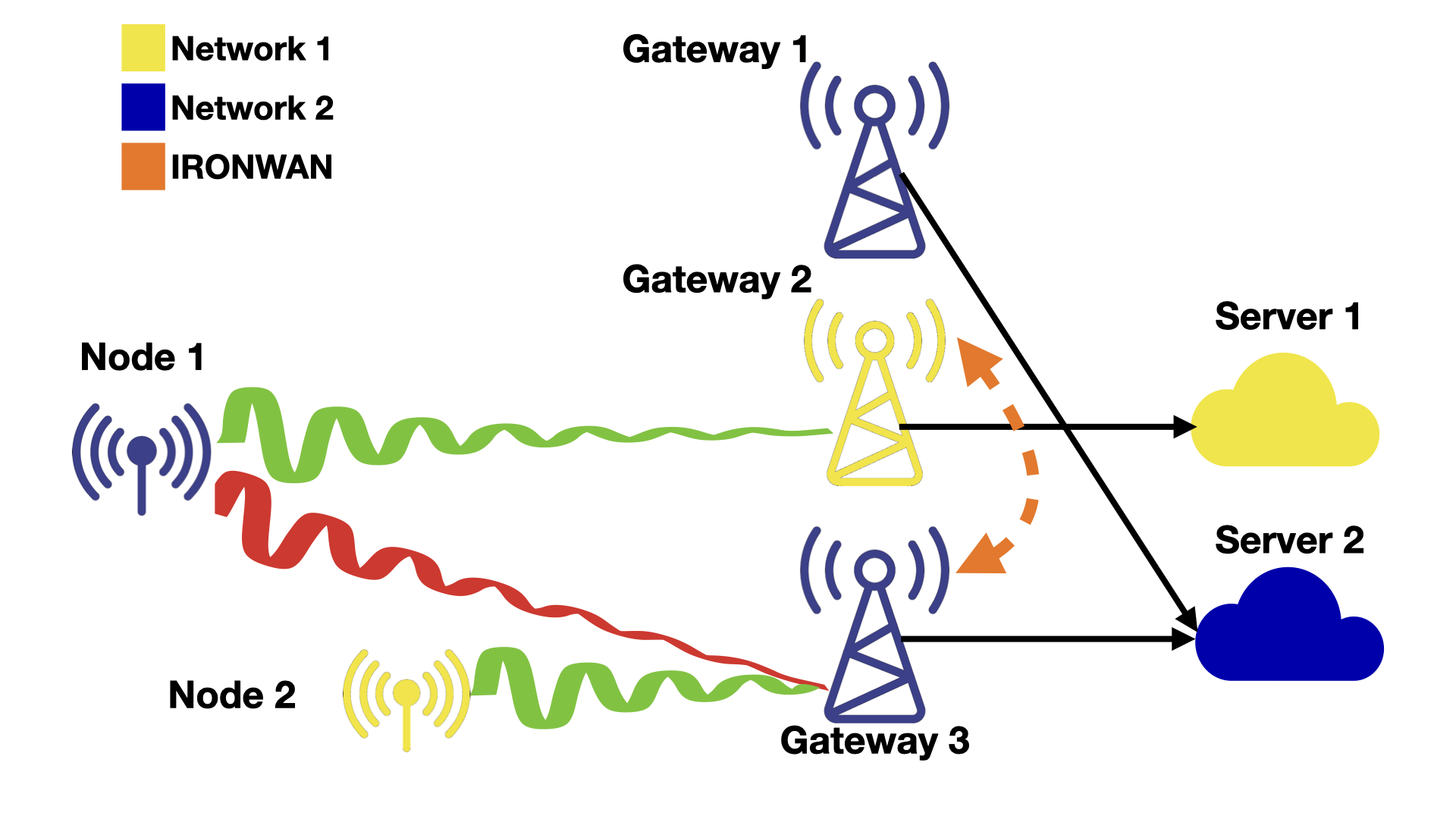}
  \caption{\AlgoName creates a wireless overlay network to link overlapping
  networks (networks 1 and 2) to exchange uplink and downlink transmissions. \AlgoName replays failed messages for uplink collisions(red colour) and can use the link for acknowledgements}
  \label{fig:teaser}
  \vspace{-5mm}
\end{figure}

In this paper, we propose \textit{\textbf{I}ncreasing \textbf{R}eliability of
\textbf{O}verlapping \textbf{N}etworks in LoRa\textbf{WAN}} (\AlgoName), a solution to the problems caused by overlapping LoRaWAN networks. It is a software approach that requires no additional backhaul networks, internet access or cloud services and can be readily deployed by only updating the gateways.  \AlgoName organises the gateways of multiple networks into a wireless overlay network for G2G message exchange (as shown in Fig.\ref{fig:teaser}). \AlgoName allows a gateway from one network to act as a redundant receiver for gateways of another network without revealing their encryption keys. Gateways operate \AlgoName can receive their uplink messages even when they do not share the same network server (i.e. they belong to different networks).   \AlgoName also enables gateways in the overlay network to share their downlink capacity. Gateways that have exhausted their allocated duty cycle on the downlink channel can send downlink messages via other gateways with free capacity. Sharing uplink and downlink messages allow all overlapping networks to increase the number of unique messages they can receive per node (which implies they have more data per node). It also helps to reduce the number of retransmissions for messages requiring acknowledgements, which saves a node's energy. Since there is no need to share encryption keys, there is no need for coordination between the network owners and no chance of the associated security threats. All that is required is adding some services on the gateways and no modification to the nodes making the adoption of \AlgoName trivial and anonymous.

Gateways in \AlgoName send G2G messages only when necessary and with local information in a fully distributed manner. Our contributions in this work are summarised as below:

\begin{itemize}[leftmargin=\parindent,topsep=1.5pt] 

    \item By analysing 11-million real-world LoRaWAN messages, we propose a Real-time Message Inter-arrival Prediction (\Arr) algorithm. With \Arr, gateways can adaptively predict message-arrival times from hundreds if not thousands of uplink nodes with only $\text{O}(n)$ computation and memory overhead. Here $n$ is a user-defined parameter (i.e. window size) that is typically small and independent of network and message sizes. Our evaluation shows \Arr can achieve more than 99\% accuracy in both precision and recall at the same time.  
    
    \item To minimise communication interference caused by the G2G communications in \AlgoName, we propose \IP. \IP exploits reinforcement-learning techniques to predict the behaviours of other nodes to avoid message collisions during G2G communication in a fully distributed manner. Our evaluation results demonstrate that \IP reduces messages collisions for G2G communications from $16\%-39\%$ to $7\%-13\%$ given different networks loads.
    
    \item We test \AlgoName (with \Arr and \IP) with a trace-driven simulation consisting of 1,000 nodes and 6-10 gateways against original LoRaWAN, FLIP \cite{FLIP} and an optimised centralised wired approach using OMNet++. Our experimental results show that \AlgoName improves packet delivery ratio (PDR) up to 28\% and reduces message re-transmissions by 50\% compared to the original LoRaWAN.  Compared to FLIP with hardware backhaul between gateways, \AlgoName shows comparable performance improvement and outperforms FLIP with more than 8 gateways. We also implemented and tested \AlgoName with a $10$-node test-bed to demonstrate its practicality.    
\end{itemize} 

We organise the paper as follows. We present preliminaries and background in Sec. \ref{sec:background}. We present an overview of \AlgoName in Sec. \ref{sec:overview}. We then describe the detailed design of \Arr and \IP in Sec. \ref{sec:interarrival} and \ref{sec:reinforcement}. We present the experimental results in Sec. \ref{sec:evaluation}, and then we round out our discussion with sections on the limitations and potential extensions of our approach \ref{sec:limitations}, related work \ref{sec:related}, and a
conclusion \ref{sec:conclusion}.

\section{Background and Preliminaries}
\label{sec:background}

In this section, we describe the current LoRaWAN architecture and the problems with overlapping networks.

\subsection{LoRaWAN Architecture Overview}
\label{subsec:lorawan_architecture}
LoRaWAN operates a 3-level architecture, consisting of \textit{nodes}, \textit{gateways} and \textit{servers}. They are (as shown in Fig.\ref{fig:teaser}):\\ 
\textbf{Nodes} are devices responsible for sensing and transmitting data to servers via gateways using LoRaWAN\cite{LoRaWAN}. These nodes use the ALOHA channel access scheme. They must implement Class-A functionality, where nodes open two receive windows(1 and 2 seconds) after transmission to receive acknowledgements for their messages. If messages are not acknowledged, nodes can retransmit these messages. The number of retransmissions and the algorithm are user-defined. \\
\textbf{Gateways} are the bridges between servers and nodes responsible for wireless communication with nodes. They receive messages via uplink channels and send control and acknowledgement messages via downlink channels. The owner of the gateway chooses to which server to forward all messages. In LoRaWAN, the gateways forward \textit{all} messages they receive to the servers even if the messages belong to other networks. \\
\textbf{Servers} are the final destination for the data collected by the nodes. Servers discard messages belonging to unknown nodes (other networks) as they are unwanted and cannot be decoded. Once messages are successfully delivered, servers select the gateway with the best link conditions with the originating node (e.g. highest RSSI, SNR) to send acknowledgements when necessary. 

\subsection{Impact of overlapping networks}
\label{subsec:exploiting_others_gateways}

\begin{figure}[ht!]
\begin{subfigure}{0.48\textwidth}
  \centering
    \includegraphics[width=\textwidth]{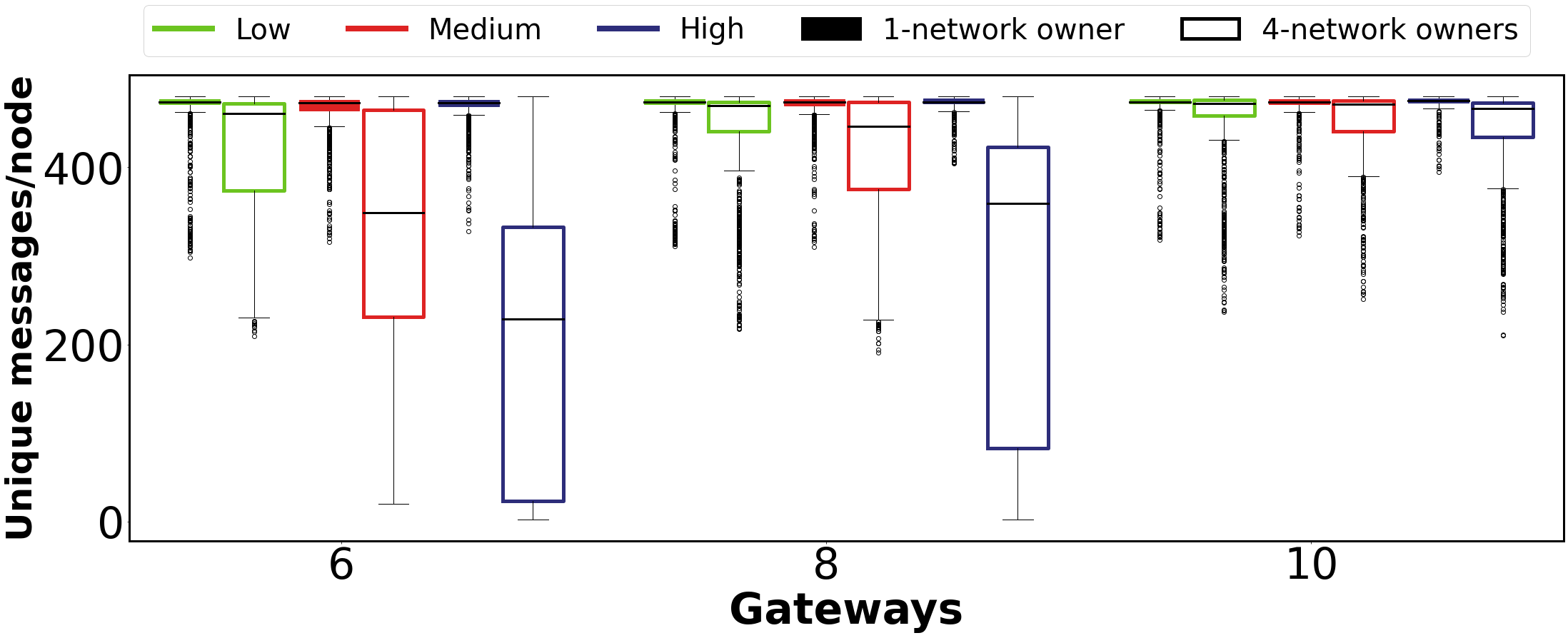}
  \caption{\label{fig:partitioning_effect_goodput_2}}
\end{subfigure}
\newline
\begin{subfigure}{0.48\textwidth}
  \centering
    \includegraphics[width=\textwidth]{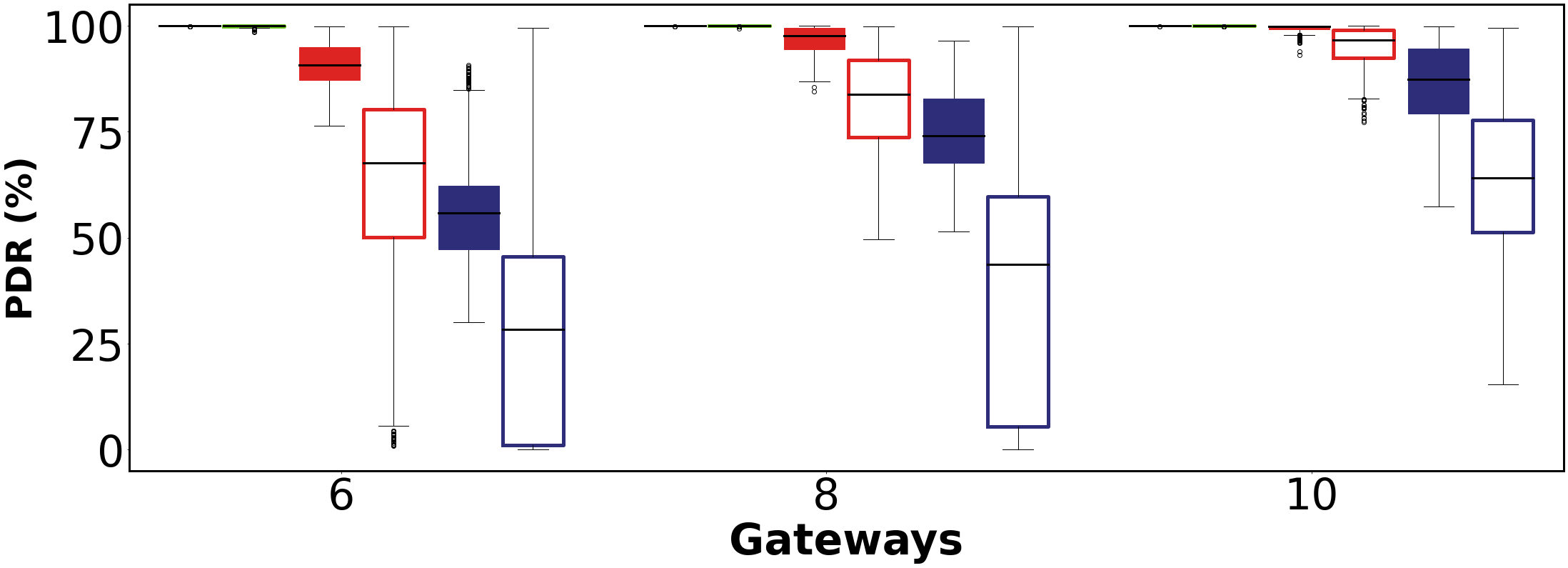}
  \caption{\label{fig:partitioning_effect_pdr_2}}
\end{subfigure}
\caption{Effect of partitioning in a LoRaWAN network}
\label{fig:partitioning}
\end{figure}

To study the effect of multiple overlapping networks on the packet delivery ratio (PDR) and the number of unique uplink messages per node received at the server (for a detailed description, see Sec.\ref{sec:evaluation}), we simulate $1000$ nodes generating a message every $3$ minutes.
Fig.\ref{fig:partitioning}  shows the results for a network with 1 owner and 4 overlapping networks with 4 owners with the same network topology. We test the networks under three levels of message acknowledgement requirements  (i.e. low:10\%, medium:50\% and high:100\%). At all three levels, the 1-owner network outperforms the networks with 4-owners. Given networks consisting of 6 gateways, the number of unique messages received per node and PDR of the 1-owner network is 65\% and 50\% higher than the network consisting of 4 different stakeholders without gateway sharing. Our analysis shows that networks with 4-owners suffer from extensive message collisions caused by message retransmissions or lack of available duty-cycle to acknowledge these messages. This work aims to introduce the performance advantages of 1-owner networks to networks with multiple owners by sharing gateway resources between multiple owners.

\section{\AlgoName Overview}
\label{sec:overview}

\AlgoName is a G2G communication system that solves the message loss and downlink duty-cycle exhaustion problems caused by overlapping LoRaWAN networks. It does this by enabling gateways to coordinate to find lost uplink messages and share downlink capacity. Importantly, \AlgoName schedules G2G communication to minimise interference to
all of the networks. \AlgoName runs as an add-on module on LoRaWAN gateways, is fully compatible with the LoRaWAN specifications, does not require any backhaul networks, does not incur usage costs, and is readily deployable with a software update on the gateways. This section presents an overview of \AlgoName's architecture and the four sub-modules forming \AlgoName. We then provide an operational overview describing the interactions between modules and the objectives and
challenges for \AlgoName.

\subsection{\AlgoName Architecture}

\AlgoName consists of four sub-modules: a manager, a cache and two learning algorithms that run on every \AlgoName-enabled gateway as shown in Fig.\ref{fig:overview}. They are: 

\begin{figure}[ht!]
    \centering         
    \includegraphics[width=0.48\textwidth]{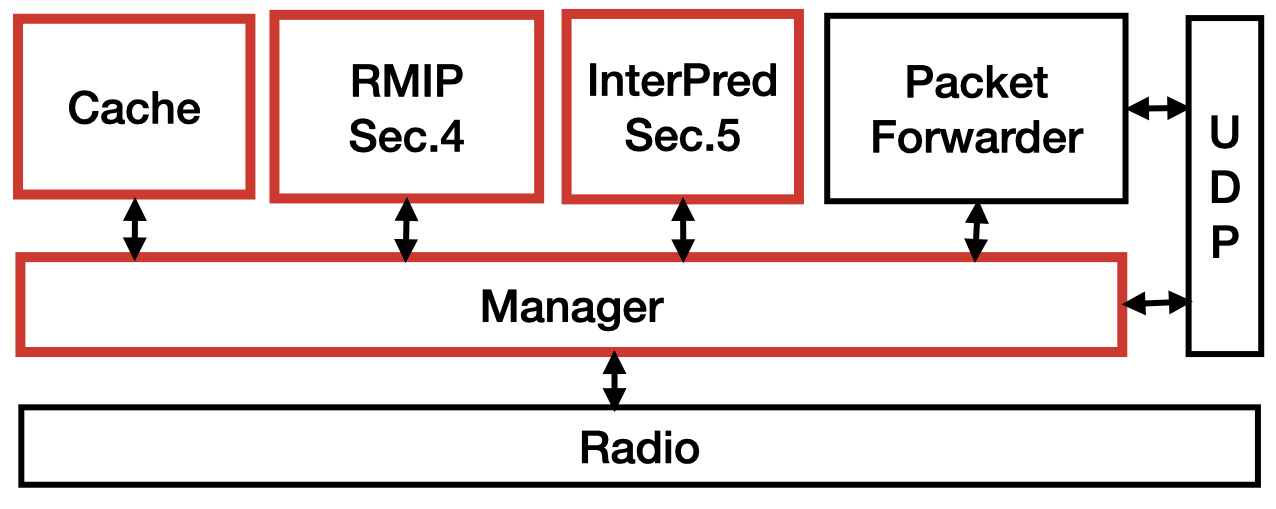}
    \caption{Interaction between LoRa radio, packet forwarder and \AlgoName's modules (shown in red)}
    \label{fig:overview}
    \vspace{-5mm}
\end{figure}

\begin{enumerate}[leftmargin=\parindent,topsep=1.5pt]
    \item \textbf{Manager module} facilitates gateway communication. This module is responsible for scheduling and orchestrating access to the wireless channel for a gateway, tracking channel usage, querying \Arr and \IP when needed and handling incoming G2G messages from other gateways. The Manager module is also responsible for requesting lost uplink messages (by querying \Arr), handing over downlink messages and ensuring that any G2G communication causes minimal interference (by querying \IP). \AlgoName also introduces new G2G messages that are not defined in standard LoRaWAN, and so the manager module is also responsible for prioritising and scheduling these messages. It also forwards messages received from the radio layer to the appropriate modules, including the Packet Forwarder (part of the LoRaWAN specification). No encryption keys need to be shared by the manager modules of different gateways. 
    
    \item \textbf{Caching Module} caches the latest data packages received from nearby nodes within the communication range. Upon receiving data requests (for lost messages) from other gateways, the manager module accesses this cache to find the requested messages. The data stored in this cache are deleted after a user-defined time to limit memory consumption.

    \item \textbf{Real-time Message Inter-Arrival Predictor (\Arr):} estimates the message-arrival time for every node within the reception range. If an
    the expected message does not arrive in time, it triggers the manager module to request the lost message from gateways belonging to other networks, or the manager module can poll this module to get information about lost messages, see Sec.\ref{sec:interarrival}. 
    
    \item \textbf{Interference Predictor (\IP):} predicts if a G2G communication can cause interference through training a value-based reinforcement-learning agent which learns channel usage in real-time. By querying this module, the manager can choose time-slots and channels for G2G communications without causing interference to N2G communication, see Sec.\ref{sec:reinforcement}.
\end{enumerate}

\subsection{Operation Overview}
\label{subsec:operationoverview}
Every gateway has to perform three operations alongside standard LoRaWAN functionality to be \AlgoName-compliant. The first is to duplicate and forward the received messages correctly. The second is to request for failed uplink messages if they haven't arrived. The third is to hand over downlink messages if a gateway has no downlink capacity. These operations are described in detail below.

\begin{figure*}
\begin{subfigure}{.33\textwidth}
  \centering
    \includegraphics[width=\textwidth]{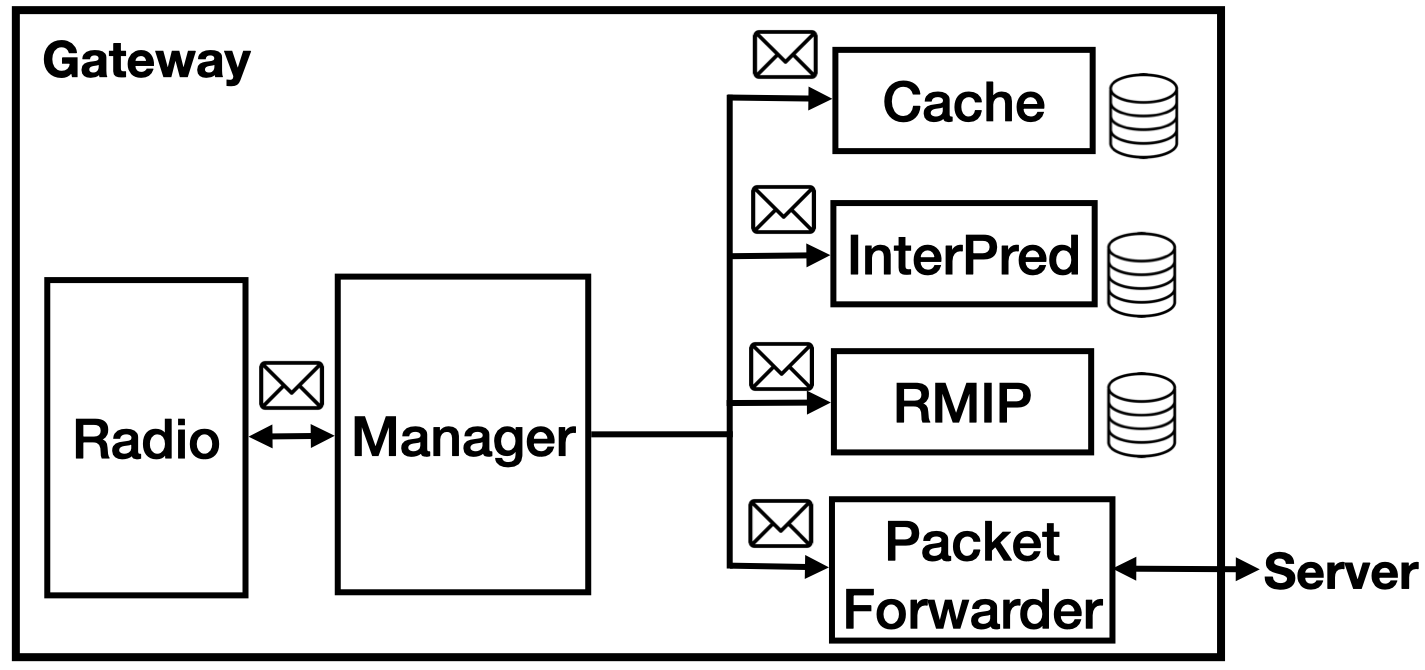}
  \caption{\label{fig:uplink}}
\end{subfigure}%
\begin{subfigure}{.33\textwidth}
  \centering
    \includegraphics[width=\textwidth]{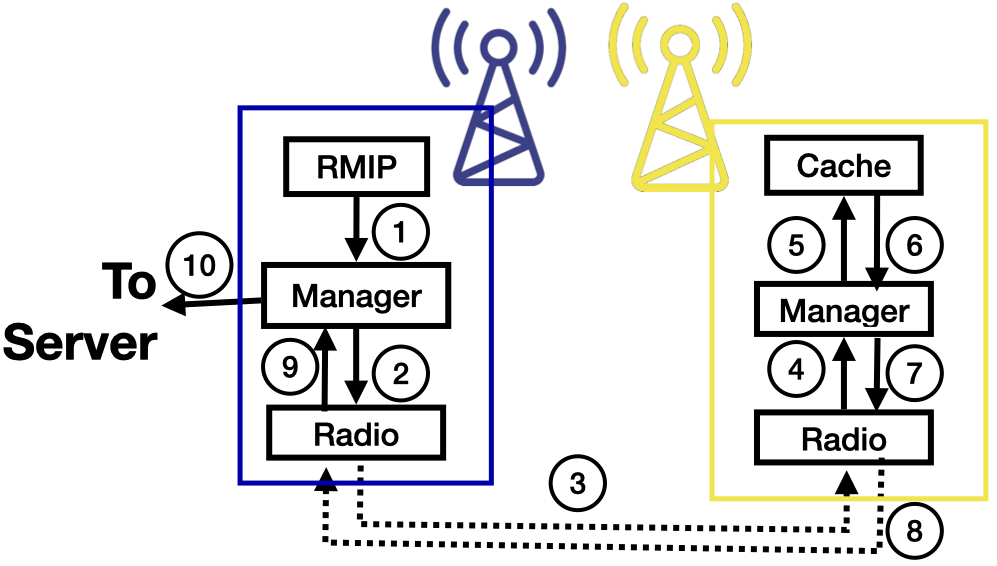}
  \caption{\label{fig:uplinksolution}}
  
\end{subfigure}
\begin{subfigure}{.33\textwidth}
  \centering
    \includegraphics[width=\textwidth]{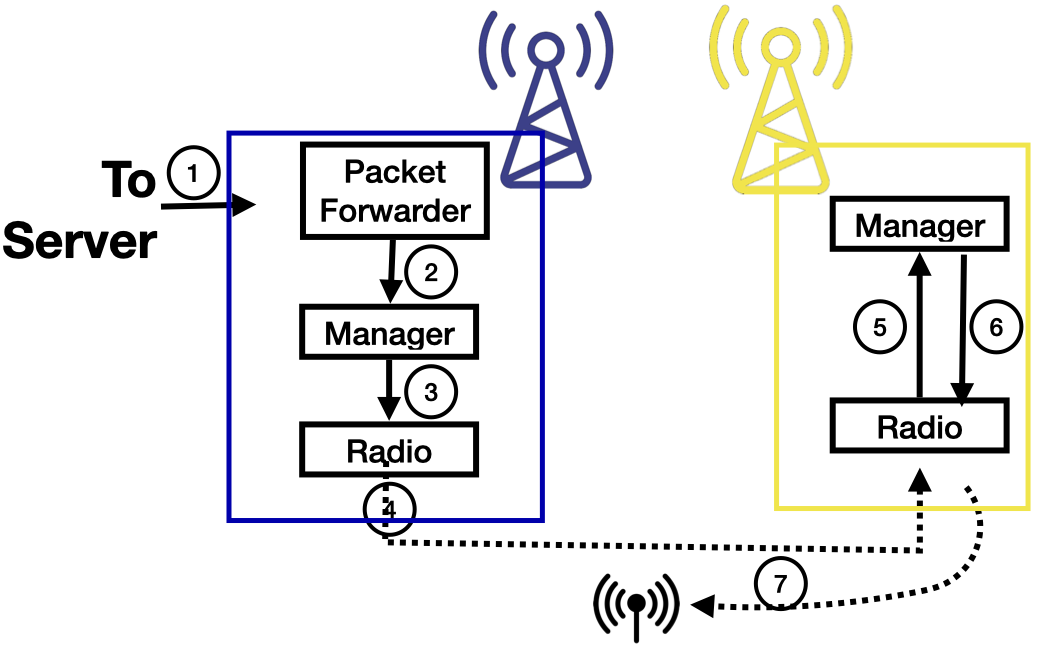}
    \caption{\label{fig:downlinksolution}}
\end{subfigure}
\vspace{-3mm}
\caption{(a) All received messages are replicated by manager and
forwarded to all modules, (b) \AlgoName's G2G communication for a gateway to
request a lost uplink message from a neighbour gateway. (c) \AlgoName's G2G communication for
a gateway with no remaining downlink duty-cycle to
request that a neighbour gateway send a downlink message.}
\vspace{-5mm}
\end{figure*}
\textbf{Received Messages.} As shown in Fig.\ref{fig:uplink}, the manager module forwards all valid LoRaWAN messages to the cache module, \Arr module, \IP module and the packet forwarder. The packet forwarder sends the message to the server as per the LoRaWAN specification.

\textbf{Request for a lost uplink message due to uplink packet collisions.}
As described in Fig.\ref{fig:uplinksolution}, a gateway's \Arr module
indicates to the manager that an uplink message from a node was lost \textcircled{1}. The manager then sends a \textit{Request for uplink message} with the node's address and last received message ID \added[]{to radio \textcircled{2} which in-turn broadcasts it using LoRa} \textcircled{3}. The manager module of the receiving gateway \textcircled{4}, queries its caching module for a newer message than the message ID for that particular node \textcircled{5}. If there is such a message\textcircled{6}, the gateway responds to the requesting gateway with a
\textit{Rebroadcast an uplink message} \added[]{to radio \textcircled{7} that then broadcasts it using LoRa \textcircled{8}}. These messages are received by the gateway\textcircled{9} and forwarded to the server\textcircled{10}. Essentially, the gateways act as store-and-replay intermediaries, and this increases overall network throughput.

\textbf{Request to send a downlink message due to duty-cycle exhaustion on downlink channels.} 
As shown in Fig.\ref{fig:downlinksolution}, \AlgoName enables a gateway with
no remaining downlink duty-cycle to handover the transmission of downlink
messages to other gateways. Whenever the manager receives a message from packet forwarder\textcircled{1}, and it cannot transmit it, the requesting gateway's \added[]{packet forwards it to manager\textcircled{2} that } encapsulates
the packet in a \textit{Request to forward Downlink} message and sends it \added[]{to radio\textcircled{3} that broadcasts it using LoRa \textcircled{4}}. If a gateway belonging to another network receives
this message, it checks its cache to verify if it has received a message from
that particular node in the last two seconds\added[]{\textcircled{5}}. If the gateway has received a
message from that node, it gets the time of when to transmit the message from
its cache, and it schedules\textcircled{6} and transmits a \textit{Neighbour Downlink
Message}\textcircled{7}. 

\subsection{Objective and Challenges}
\label{subsec:challenges}
\AlgoName solves the problem caused by overlapping networks by exchanging failed uplink and downlink messages between gateways. \AlgoName uses LoRaWAN for G2G communication.
LoRaWAN operates in \deleted[]{in} the sub-GHz unlicensed band, which is subject to a 1\% 
duty-cycle on band 0 (868.0-868.8 MHz for both uplink and downlink) and a 10\% 
duty-cycle on band 1 (869.40-869.65 MHz for downlink) in the EU. As the access to the channel is duty-cycle limited, this makes communication a scarce resource. So message transmissions need to be planned and scheduled to use the spectrum efficiently. Gateways are also resource-constrained devices (e.g. raspberry pi) and may connect to hundreds if not thousands of nodes simultaneously. It is essential to ensure all of the solutions are lightweight and can scale to large networks. This creates two new challenges: 
\\
\textbf{1. Deciding when to hunt for missing messages.} We aim to solve the problem of uplink packet collisions by requesting the gateways of other networks to retransmit a message that failed to arrive at its own network as is described in Subsec.~\ref{subsec:operationoverview}.
Gateways have to accurately estimate when to expect message arrivals for every node in their communication range. To understand if this could be estimated, we analyse $11$-million real-world LoRaWAN messages to see if there was an observable trend, using the LoED dataset \cite{bhatia2020loed}. \added[]{LoED is a real-world dataset consisting of LoRaWAN messages collected by passively listening at $9$ gateways deployed in London, representing a dense urban environment.} LoED's data showed that message inter-arrival times per node are relatively consistent. 56\% of the nodes in the dataset sent messages periodically, 5\% of these nodes changed their transmission period at some point because of application layer changes or the reassignment of node ids, and only 4\% of the nodes which transmit messages periodically require acknowledgements. This observation is in line with the results in cite{Choi2020}. Their results show that around 65\% of the nodes transmit with intervals that have less than 10 seconds of standard deviation. These results indicate that gateways can predict the message arrival time for nodes that send messages periodically. An additional solution is also required to cope with changes in inter-arrival periods and errors from undefined behaviours that could deviate from our predictions.
\\
\textbf{2. G2G communication scheduling.} Whenever a G2G communication occurs to exchange uplink or downlink messages, it can only occur on band 0 as gateways only listen on that band. G2G messages will interfere with node-to-gateway (N2G) communication on band 0. A gateway can choose from several channels and times on this band to transmit a packet. If it transmits at the wrong time on the wrong channel, it will interfere with an N2G message, leading to a lost packet, triggering retransmission. We need an effective, lightweight solution to predict and avoid interference so gateways can schedule G2G transmissions that cause the least interference. The solution needs to consider the changes in the environment, like varying transmission periods and network conditions that affect communication parameters. 

In this section, we have described \AlgoName's components, how they interact, and the communication protocol. In the next two sections, we will dive deeper into the design and evaluation of \Arr and \IP modules, which are solutions for the objectives described in Subsec.\ref{subsec:challenges}.
\section{Estimating Message Arrival with \Arr}
\label{sec:interarrival}

In this section we describe how the \Arr module 
estimates the message inter-arrival time for every node. 
\Arr uses techniques from streaming and statistical methods. The 
computation and memory overhead of \Arr are both $O(n)$, where $n$ is a buffer size that determines how fast \Arr responds to changes in inter-arrival 
times. 
A gateway only needs $(n + 1) * 4$ bytes (assuming 32-bit floats) per node and can predict message arrivals from hundreds if not thousands of nodes using low memory and computation.

\subsection{Problem Statement}
\begin{figure}[ht!]
    \includegraphics[width=0.48\textwidth]{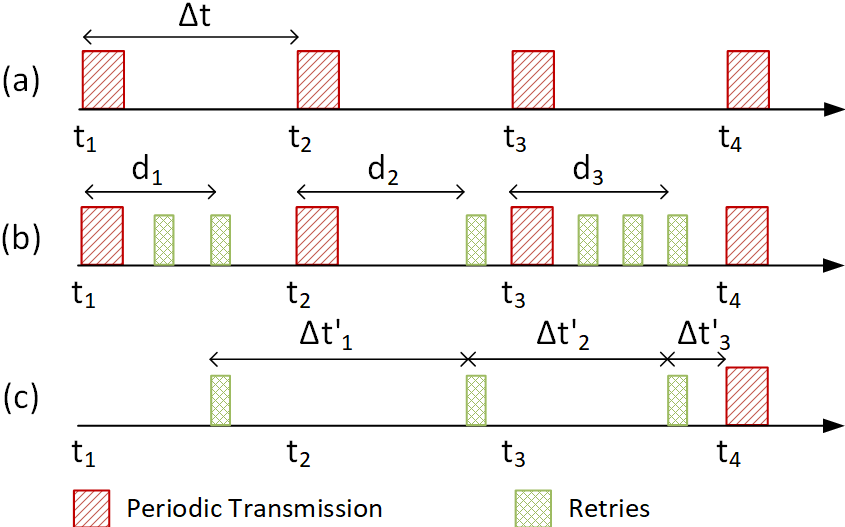}
    \caption{Examples of: (a) uplink messages sent and received by a node and the gateway, respectively, without retries. (b) uplink messages sent by a node with retries. (c) uplink messages received at the gateway with retries.}
    \label{fig:inter-arrival-time-example}
    \vspace{-4mm}
\end{figure}

Assume a gateway with a set of nodes in its reception range. The nodes transmit messages periodically with an inter-arrival time $\Delta t$.
We show in Fig.\ref{fig:inter-arrival-time-example}(a), when every uplink message successfully arrives at a gateway on the first try (i.e. no retransmissions),
the messages arrival time can be seen as a time series $\langle t_1,t_2,\cdots,t_l\rangle$, where $\{1,2,\cdots,l\}$ are monotonically increasing message IDs. In Fig.\ref{fig:inter-arrival-time-example}(b) we see that for messages that require acknowledgements nodes will try to retransmit the message when a message delivery fails before the next message is generated. For messages that do not need acknowledgements, this message is lost. The actual arrival time at the gateway, shown in Fig.\ref{fig:inter-arrival-time-example}(c), becomes $\langle t_1+d_1,t_2+d_2, \cdots,t_m+d_m\rangle$, where $\{d_1, d_2, \cdots, d_l\}$ are random numbers in range $[0,\Delta t)$ denoting the delays incurred because of these retransmissions. As messages may be missing and these delays are unknown to the gateway, message delivery intervals may vary and become $\langle \Delta t'_1,\Delta t'_2, \cdots,\Delta t'_m\rangle$.

Our objective is, given the \textit{observed} message inter-arrival times $\langle \Delta t'_1,\Delta t'_2, \cdots,\Delta t'_m\rangle$, to predict the \textit{actual} message arrival times $\langle t_1,t_2, \cdots,t_l\rangle$ at the gateway.
We further break down the problem into three sub-problems listed below:
\begin{itemize}[leftmargin=\parindent,topsep=1.5pt]
    \item \textbf{Inter-arrival time prediction:} The first task is to predict the actual inter-arrival time $\Delta t$ given $\langle \Delta t'_1,\Delta t'_2,\cdots,\Delta t'_l\rangle$.
    \item \textbf{Reference-anchor prediction:} The second task is to find a reference point from where this inter-arrival time is valid. To predict $\langle t_1,t_2, \cdots,t_l\rangle$ from inter-arrival time $\Delta t$, we find a reference time point $t_\varnothing$ where a message arrives with \textit{zero} retries.
    \item \textbf{Change detection in inter-arrival time:} The inter-arrival time $\Delta t$ may change overtime. Our predictions based on previous observations may be skewed if the gateway is not aware of these changes. Consequently, a gateway needs to detect changes and adapt accordingly given $\langle \Delta t'_1,\Delta t'_2,\cdots,\Delta t'_l\rangle$. 
\end{itemize}

It is worth noting that, each LoRaWAN gateway may connect to thousands of LoRaWAN nodes and it is essential to minimise the extra overhead introduced by \Arr.

\subsection{Estimating $\Delta t'$ for Missing Messages}
Before starting the prediction algorithm, \Arr first checks if each data point in $\langle \Delta t'_1,\Delta t'_2, \cdots,\Delta t'_m\rangle$ is computed from messages with two consecutive IDs (e.g. $\Delta t'_1 = t'_2-t'_1$). 
In practice, some messages may go missing and never reach the gateway (delay $d_i>\Delta t$) due to interference.
The Inter-arrival time computed from a message stream with lost messages can significantly deviate from the ground truth. %
For example, when a gateway only receives every alternate message, the inter-arrival estimation can be two times larger than the ground truth.
To deal with missing messages \Arr fills in these missing $\Delta t'_i$ using the equation below:
\[\Delta t'_i = \frac{t'_{l+m}-t'_{l}}{m} \forall l<i<l+m \]
where $l$ and $l+m$ denotes the message ID of the recent and last received messages, and $m$ denotes the number of missing messages.

\subsection{Inter-Arrival Time Prediction}
\label{subsec:RMIP_S1_inter-arrival}

Once the missing inter-arrival times are estimated, the next task is to estimate $\Delta t$ given $\langle \Delta t'_1,\Delta t'_2,\cdots,\Delta t'_l\rangle$.
To do so, we have a closer look at the message inter-arrival time observed in real-world scenarios. From example shown in Fig.\ref{fig:inter-arrival-time-example}, we observe the property:

\begin{equation}\label{eq:mean_error}
(l-1) \Delta t < \Sigma_{i=1}^l \Delta t'_i < (l+1) \Delta t,
\end{equation}
where $l$ is the number of messages expected.
From Eq.\eqref{eq:mean_error}, we know that the error term is bounded by $\pm 1/l$ when using the \textit{mean} as our estimator.
However, using the mean has several drawbacks. First, this error term is not equal to zero unless the first and last messages both arrive without delay. This is a problem in practical scenarios where delays due to collisions, retransmissions or clock jitter are common. The estimation may deviate from the actual inter-arrival time.
Second, the mean value is more sensitive to outliers, and a large delay may significantly skew the estimation. We use the \textit{median} as the estimator, which is more robust against the problems mentioned above.

The median may still deviate from actual intervals when there are no \textit{exact} inter-arrival times in observed samples. We overcome this problem by performing a statistical test on the inter-arrival estimate.
Due to the small sample size ($n$) ($n=10$ in our implementation), \Arr uses student's t-test\cite{de2013using} to test if we should use the computed median value as $\Delta t$.
T-test produces a p-value (with regards to its degree of freedom) which we test to see if we can reject the null hypothesis (i.e. median is not valid). 
We adopt a more restrictive p-value ($0.703$ representing a 50\% two-side quantile when $n=10$) for our implementation. 
If the null hypothesis is rejected \Arr accepts this median as the estimated inter-arrival time ($\Delta t$), otherwise it continues to collect new inter-arrival time samples while dropping the oldest one until the null hypothesis is rejected. 

\subsection{Reference-Anchor Prediction}
\label{subsec:RMIP_S2_reference_prediction}
Our next task is to find the reference time point $t_\varnothing$ with which $\langle t_1,t_2, \cdots,t_l\rangle$ can be acquired.
\Arr uses a simple but efficient method to predict $t_\varnothing$ using the equation below:
\begin{equation}
\label{eq:beginning}
   t_\varnothing = \left\{
   \begin{array}{l l}
      t_n & t_n-t_\varnothing < n\Delta t, \\
      t_\varnothing & \text{else},
   \end{array} \right.
\end{equation}
where $n$ is the message counter values from $t_\varnothing$.
As can be seen, it resets $t_\varnothing$ when messages are received before expected arrival time.
This is valid because LoRaWAN message counter values are reset only when a new message is generated.
This allows $t_\varnothing$ to quickly converge to the \textit{first} transmission for any message of every node. %

\subsection{Change Detection in Inter-Arrival Time}
\label{subsec:RMIP_S3_change_detection}
\Arr adopts the event trigger technique that is commonly seen in stream processing to minimise memory and computational requirements and detect changes in the inter-arrival time. 
Instead of updating $\Delta t$ every time a new message is received, it compares the new inter-arrival time with $\Delta t$.
If \Arr collects $n$ \textit{consecutive} inter-arrival times $\langle \Delta t'_1,\Delta t'_2,\cdots,\Delta t'_n\rangle$, where their difference with $\Delta t$ is greater than a given threshold $e$ (i.e. $|\Delta t - \Delta t'_i| > e, \forall i\in \{1,2,\cdots, n\}$), it assumes that a node's transmission interval has changed. \Arr then recomputes the $\Delta t$ and $t_\varnothing$, with the algorithms presented in Subsecs.\ref{subsec:RMIP_S1_inter-arrival} and \ref{subsec:RMIP_S2_reference_prediction}, respectively.

\begin{figure}[ht!]
    \begin{subfigure}{0.48\textwidth}
      \centering
        \includegraphics[width=\textwidth]{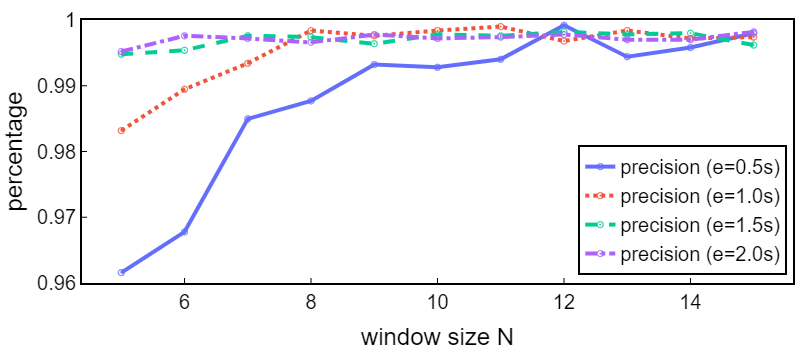}
        \caption{Precision}
    \end{subfigure}
    \newline
    \begin{subfigure}{0.48\textwidth}
      \centering
        \includegraphics[width=\textwidth]{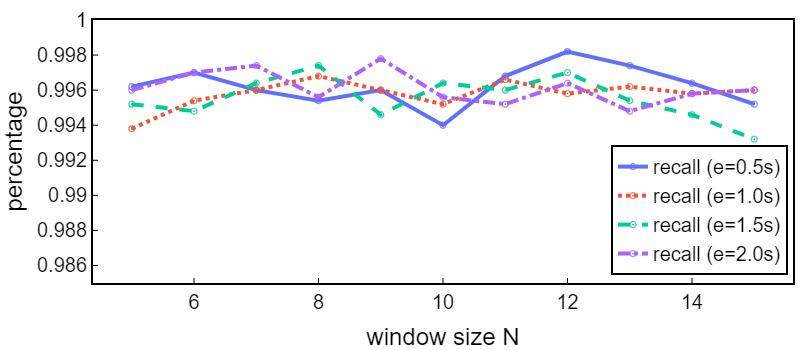}
        \caption{Recall}
    \end{subfigure}
    \caption{The precision and recall of the change detection in RMIP given given different window size $n$ (5-15) and error threshold $e$ (0.5s-2.0s)}
    \label{fig:RMIP_parameter}
    \vspace{-6mm}
\end{figure}

To better understand the impact of $n$ and $e$ on \Arr, we run an experiment on a real-world dataset, LoED\cite{bhatia2020loed}. As the ground-truth changes are not available in this dataset, changes were simulated by concatenating inter-arrival timeseries of different nodes (to simulate artificial inter-arrival changes). The results are shown in Fig.\ref{fig:RMIP_parameter}. As can be seen, \Arr captures changes with a very high accuracy. Precision and Recall are more than 96\% in all of our experiments. There is no discernible changes in Recall with respect to both $n$ and $e$. Precision improves with larger $n$ and $e$; however, the difference become negligible when $e\geq 1.5$s.
In our experiments, we choose $n=10$ and $e=1$s as default. This is because LoRaWAN devices open two receive windows 0.9-1.1 and 1.9-2.1 seconds for acknowledgements from gateways as presented in Sec.\ref{subsec:lorawan_architecture}. 

\section{Interference Prediction with \IP}
\label{sec:reinforcement} 
In this section we describe \IP. 
Each gateway using \AlgoName has an \IP agent, when requested, determining a channel and timeslot for G2G communication that will not interfere with N2G communication.
\subsection{\IP Overview}
\IP trains an \textit{interference predicting agent} on each gateway
by overhearing the network traffic that reflect the communication usage of the neighbouring gateways and nodes. 
The agent learns an interference model to decide the channel and timeslot for G2G communication that will not interfere with other communication.
To formalise LoRaWAN interference prediction as a \textit{reinforcement learning} problem, \IP introduces novel definitions of state, action, policy, and reward, and uses the value-based reinforcement learning method, State-Action-Reward-State-Action (SARSA)\cite{sutton2018reinforcement}, to train the gateway agent.
We use SARSA because of its merits in embedded applications (e.g., fast, efficient, and no pre-trained model requirement).
It is not trivial to use SARSA for agent training to predict wireless communication. Most wireless communication systems (including but not limited to LoRaWAN) are \textit{half-duplex}, i.e., messages can be either sent or received on a channel but not both simultaneously. A half-duplex system does not have the immediate feedback about its success or failure required by SARSA.
To address this issue \IP trains the agent based on \textit{pseudo actions} (discussed in Sec.\ref{subsec:pseudoaction}).

\begin{figure}[ht!]
    \centering         
    \includegraphics[width=0.48\textwidth]{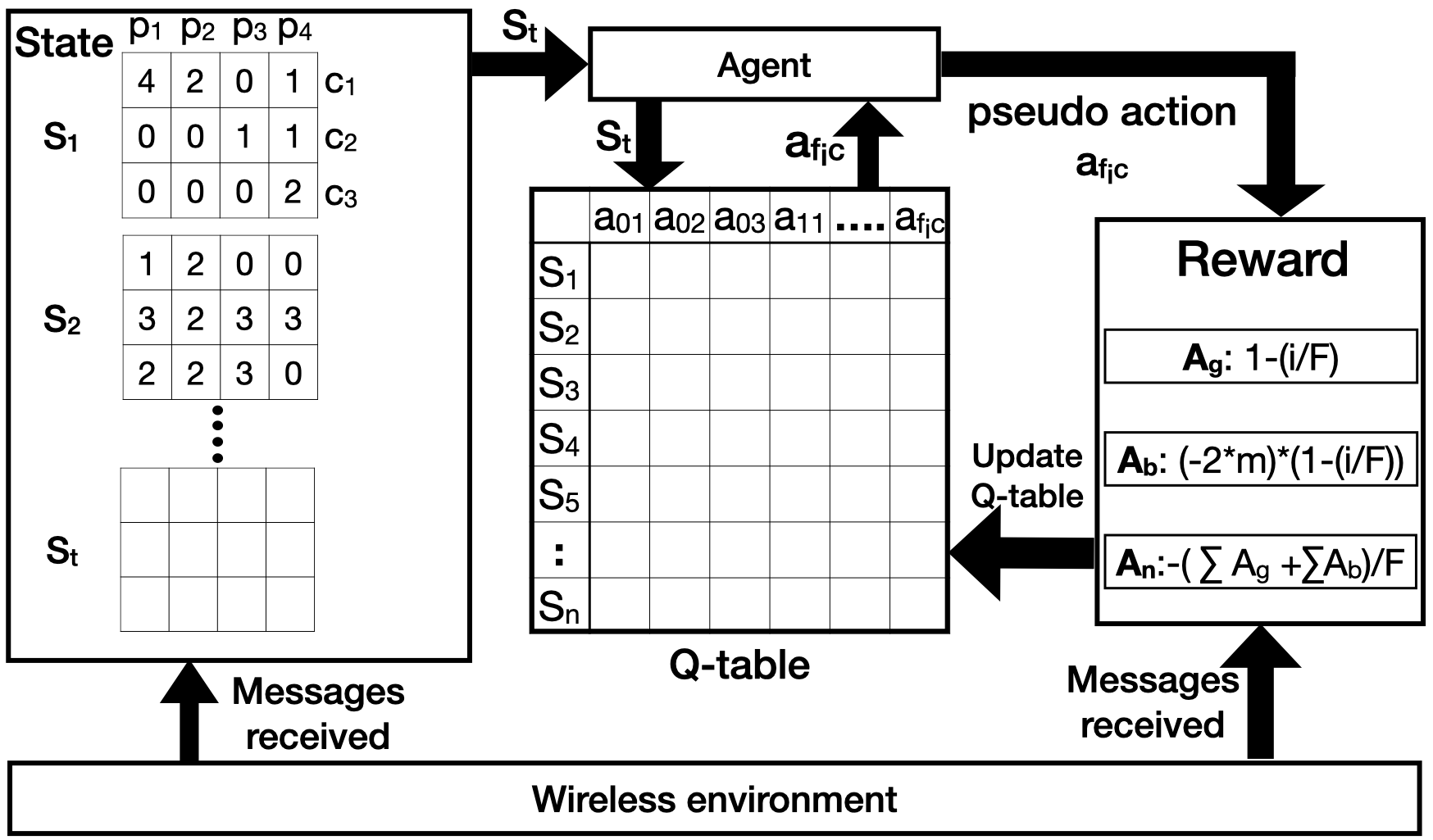}
    \caption{\IP overview}
    \label{fig:rloverview}
    \vspace{-5mm}
\end{figure}

The workflow of \IP is illustrated in Fig.\ref{fig:rloverview}.
An \IP agent on a specific gateway chooses an \textit{action} (i.e., decides the channel and timeslot inducing the minimal interference with N2G communication, and conducts G2G communication correspondingly) given a specific \textit{state} (i.e., the network status defined by spectrum usage information during a past period of time).
Since the optimal acting \textit{policy} (i.e., best actions to take in different states) is unknown after the system initialisation, the agent continuously interacts with its \textit{wireless environment}, and iteratively optimises its policy according to the environment \textit{reward} determined by the interference caused by actions taken.
We present detailed definitions as follows.

\subsection{State, Action, Policy, and Reward}

Formally, we assume that \AlgoName operates in discrete timeslots $t \in {1,2, \cdots}$ of $0.1$ seconds.
\IP stores the communication spectrum usage information of up to $P$ past slots.
When a gateway needs to conduct a G2G communication, it uses the information collected in these $P$ slots to choose one slot from future $F$ slots on one of $C$ channels to communicate with the least possibility of causing interference with other N2G communication.
The gateways have no information about interference at other gateways, and they can only make locally optimal decisions.

\subsubsection{\textbf{State.}}

At a specific timeslot, we define the current \textit{State} $\boldsymbol{S}$ as a matrix of information about message received on each of $C$ channels in each of past $P$ timeslots (see Fig.\ref{fig:rloverview}).
Each timeslot is labelled as $p_i$, $i \in \{1, 2, ..., P\}$, and each channel is labelled as $c_j$, $j \in \{1, 2, ..., C\}$ .
$p_1$ and $p_P$ represent the oldest and the current timeslots in $\boldsymbol{S}$, respectively.
Each state matrix element $s_{{p_i}{c_j}}$ contains information about the number of messages received on $c_j$ at $p_i$.
Whenever the gateway receives a message in $p_P$ on $c_j$, it extracts information about the airtime of a packet, and calculates the number of timeslots where packet reception was undergoing ($k$) by dividing airtime with slot length ($0.1$ seconds).
$s_{{p_i}{c_j}}$ is then updated as:
\begin{equation}
	\begin{array}{l l}
	s_{{p_i}{c_j}} = s_{{p_i}{c_j}} + 1, & i \in [{max(0,P-k),P]}.
	\end{array} 	
\end{equation}
We use this approach as the gateway radio only forwards a message when it has been completely received. The maximum value of $s_{{p_i}{c_j}}$ is bounded by $5$, which implies that $c_j$ is congested and not suitable for transmission.
Such a bound objectively reflects this property of LoRaWAN communication, and it helps to restrict the state space for better tractability.

\subsubsection{\textbf{Action.}}
In a specific state $\boldsymbol{S}$ at timeslot $p_P$, if G2G communication is required, the \IP agent needs to take an action.
We define an \textit{Action} $a$ as the gateway conducting a G2G transmission at a future timeslot $f_i \in \{0, p_P+1, p_P+2, \cdots, p_P+F\}$ on channel $c_j \in \{c_1, c_2, \cdots c_C\}$.
Note that the agent only decides to transmit in one of next $F$ timeslots, and $f_i = 0$ represents the case where no transmission is conducted.
Specifically, $a_{{f_i}{c_j}}$ denotes transmit at timeslot $f_i = p_P + i$ on channel $c_j$.

\subsubsection{\textbf{Policy.}}

In state $\boldsymbol{S}$, the \IP agent needs to choose an action from $((F+1) * C)$ available candidates to minimise the interference on N2G communication.
To achieve this, we use a \textit{Q-value} ($Q_{\boldsymbol{S},a}$) to denote the impact of taking action $a$ in state $S$ on the N2G communication interference. A higher Q-value represents less interference.
The \textit{Policy} of an \IP agent is a table of the Q-value of each action in each state, or a \textit{Q-table} (see Fig.\ref{fig:rloverview}).
Agents take actions in an $\epsilon$-greedy manner according to the Q-table, i.e., selects the action with the highest Q-value with a probability of $1-\epsilon$, or a random action with a probability of $\epsilon$. $\epsilon \in [0,1]$ can be adjusted for a desirable trade-off between exploitation and exploration.

Initially, all elements in the Q-table are assigned with $0$ since there is no prior information about each action's impact.
Iteratively taking actions according to its latest policy, the \IP agent updates the Q-table based on the feedback from the wireless environment.
Each element $Q_{\boldsymbol{S},a}$ is updated as follows:
\begin{equation}
Q_{\boldsymbol{S},a}^{new} = Q_{\boldsymbol{S},a} + \alpha * (reward + \gamma * Q_{\boldsymbol{S}',a'} - Q_{\boldsymbol{S},a}).
\label{eq:learning}
\end{equation}
Here, $\boldsymbol{S}'$ denotes the state transferred from $\boldsymbol{S}$ after $a$ is taken, $a'$ denotes the action taken in $\boldsymbol{S}'$ according to the current policy,
$\alpha \in [0,1]$ is the learning rate that controls the stepsize of Q-value update, and $\gamma \in [0,1]$ is the discount factor that controls the weight of future Q-value.
Both $\alpha$ and $\gamma$ are empirically tuned to improve the performance of the learnt policy under our experimental settings.
\textit{Reward} is a real value quantifying how to adjust the Q-value of each action in a state considering its impact.

\subsubsection{\textbf{Reward.}}
Intuitively, if an action causes no interference,
we call it \textit{good}, which should impart a positive reward.
Similarly, a \textit{bad} action causing interference should impart a negative reward.
Moreover, if \textit{no} G2G transmission is conducted, the agent should be punished or rewarded based on whether potential chances are missed or not.
Therefore, we have following definitions:
\begin{itemize}[leftmargin=\parindent,topsep=1.5pt]
\item \textbf{Reward of a good action ($A_g$):}
If action $a_{{f_i}{c}}$ interferes with no N2G message, its reward is:
\begin{equation}
    reward_{a_{{f_i}{c}}} = 1 - i/F.
\end{equation}
This encourages the gateway to carry out G2G communication as soon as possible.

\item \textbf{Reward of a bad action ($A_b$):}
If action $a_{{f_i}{c}}$ interferes with $m$ N2G messages, its reward is:
\begin{equation}
    reward_{a_{{f_i}{c}}} = -2 * m * (1 - i/F).
\end{equation}
This introduces a penalty to discourage the gateway from taking bad actions in the current timeslot.

\item \textbf{Reward of a no-transmission action ($A_n$):}
If action $a_{{0}{c}}$ is taken, its reward is:
\begin{equation}
    reward_{a_{{0}{c}}} = (\sum A_g + \sum A_b)/F.
\end{equation}
Here, $\sum A_g + \sum A_b$ represents the total reward that would impart by all other actions, implying whether it was a lost of transmission opportunity or a good choice to not transmit.
\end{itemize}

\subsection{Agent Training based on Pseudo Actions}
\label{subsec:pseudoaction}
The real challenge of training the \IP agent with SARSA is that, LoRaWAN communication is half-duplex. The gateways can only transmit or receive on a channel at any given point, which makes it impossible to get feedback to calculate rewards. 
To address this issue, we use \textit{pseudo actions} for training.
Once deployed on a gateway with the initial policy (i.e., all Q-values are 0), the agent first enters the \textit{training phase}, where it iteratively selects actions in different states according to its policy. However, no G2G communication is actually conducted.
In the meantime, the gateway keeps overhearing all communications on all channels, and continuously provides the agent with the wireless spectrum usage information.
With this, interference that \textbf{would have been caused} by corresponding actual actions can be inferred and used to update the Q-table according to Eq.\ref{eq:learning}.

After trained based on pseudo actions for 3 hours in a 24-hour simulation, the agent is able to start actual predictions.
When the gateway needs to transmit a G2G message the agent determines a timeslot and a channel according to its Q-table for G2G transmission. The Q-table is not updated after a real transmission.
When there is no G2G transmission request the agent continuously learns based on pseudo actions.
There may be no transmission decision that the agent can take that gives it a positive reward. In this case, the agent will take the no-transmission action, and the gateway will not send a G2G message.
This lack of a decision helps \IP to deal with network conditions of extreme overcrowding to prevent complete resource starvation.

\subsection{\IP Validation}
We implement \IP and compare it with two naive policies, i.e., \textit{random} and \textit{next-used} to validate the correctness of our method.
Three agents with different prediction methods were placed under the same virtual wireless environment (defined by the same set of simulated and real-world datasets).
They were evaluated in terms of \textit{fulfilling G2G communication requests} and \textit{preventing N2G interference}.
Only the \IP agent was required for this validation.

\subsubsection{\textbf{Comparatives.}}
We selected following two comparatives:
\begin{itemize}[leftmargin=\parindent]
    \item \textbf{Random:}
    In a state, the agent randomly chooses a timeslot and channel with a uniform probability for G2G communication.
    \item \textbf{Next-Used:}
    In a state, the agent chooses the next timeslot on a channel where there was no message in the last timeslot of that channel for G2G communication.
\end{itemize}
The next-used policy requires little storage (the number of messages sent on each channel in the last timeslot).
We selected these comparatives to demonstrate how a trivial solutions perform under real-world wireless communication conditions.

\subsubsection{\textbf{Scenarios and Metrics.}}
We tested all agents in virtual wireless scenarios defined by three simulated datasets (i.e., low, medium, and high load) and the real-world dataset, LoED\cite{bhatia2020loed}. We define load as the proportion of nodes that require acknowledgements for all messages. Traffic amounts of the medium and high load scenarios are 1.5 and 2.5 times as that of the low load scenario and that of real-world dataset are lower than the low load scenario.

In all scenarios the system parameters are set to $P = 4$, $F = 8$, and $C = 3$.
$P$ is chosen based on the available memory and the convergence time of a network.
Increasing $P$ increases the convergence time that reduces the accuracy of our system.
We set $F=8$ to give enough time to facilitate G2G communications because LoRaWAN nodes open their receive windows after a fixed period
LoRaWAN specifications require all devices to operate on at least $3$ channels, so we set $C=3$.
For \IP parameters, we set $\alpha = 0.8$, $\gamma = 0.1$, and $\epsilon = 0.2$ according to empirical studies on all datasets.

The system operate for 24 hours under each scenario. We collected the \textit{numbers of different actions} taken by each agent to quantify its ability to fulfil G2G communication requests, and the \textit{total reward} received by each agent to quantify its ability to prevent N2G interference.

\subsubsection{\textbf{Memory requirements}}
\IP has a low memory footprint so that it can run on low-resource gateways. The memory requirements for \IP depends upon $P$, $F$, $C$ and the maximum number of messages per slot. With our chosen values, the maximum number of bits required to encode the counter value of $5$ is $3$ and with $P=4$ and $C=3$, we only need $P*C*3 = 36$ bits (rounded to $5 bytes$) to encode a single state. Total number of states is $((5^P)*C)= 1875$. Assuming every action is a 32-float value, the memory needed to encode all actions of a state is $((F+1) * C) * 4 = 108 bytes$. The total memory needed is $1875 * (108 + 5) ~= 210 Kbytes$. The memory requirement can be reduced by using a 16-bit float or reducing number of future states.

\subsubsection{\textbf{Computation requirements}}
\added[]{
\IP has a low computation complexity. For each gateway, $Q_{\boldsymbol{S},a}^{new}$ is calculated and the Q-table is updated at every time slot.
The lookup and computation complexity for SARSA is $O(1)$, and is one of the primary reasons to choose a lightweight solution to run on gateways that have low resources.
}

\subsubsection{\textbf{Results.}}
\begin{figure}[ht!]
    \centering         
    \includegraphics[width=0.48\textwidth]{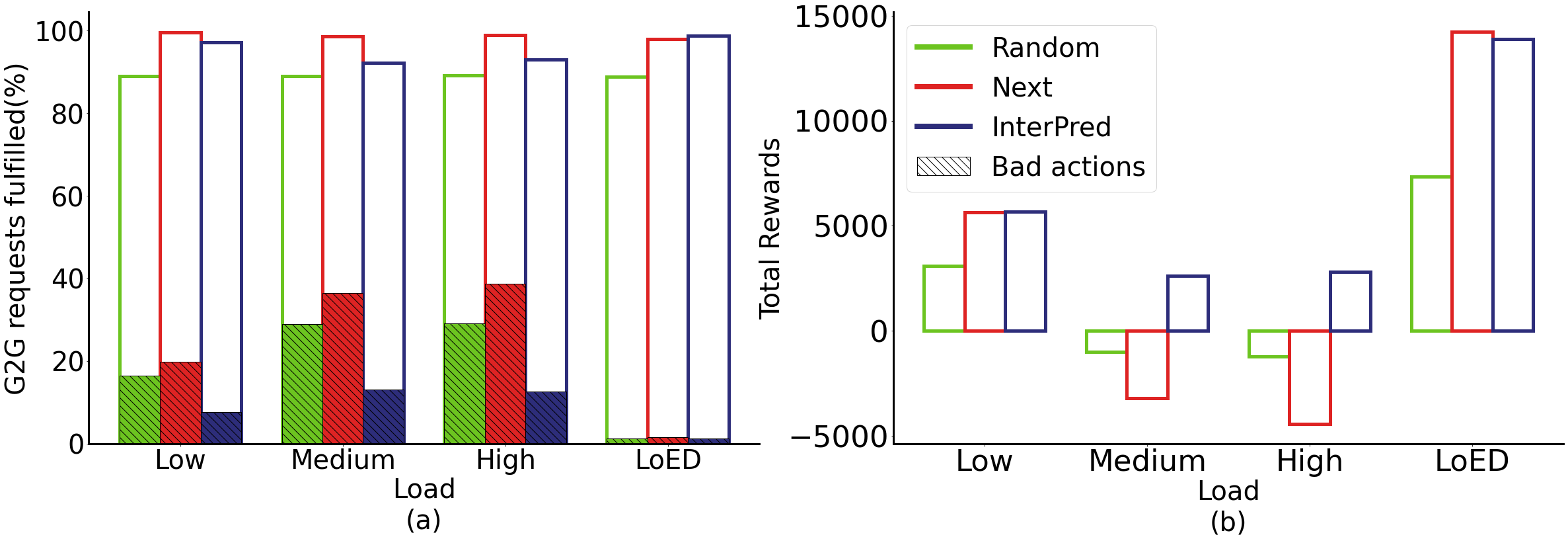}
    \caption{\IP validation results}
    \label{fig:rlvalidation}
    \vspace{-3mm}
\end{figure}

Our experimental results are illustrated in Fig.\ref{fig:rlvalidation}. \added[]{The results show that \IP performs similarly to random and next in the LoED dataset. In low-load scenario, the total rewards for \IP and next are similar, however, next has up to 2x more bad actions compared to \IP. \IP also performs better in medium and high load scenarios in terms of lower bad actions and higher total rewards.}
In scenarios defined by simulated datasets, according to Fig.\ref{fig:rlvalidation}(a), the \IP agent fulfils between 92\% and 97\% of all G2G communication requests. The ratio of bad actions to all actions 
is between 7\% and 13\%.
The random agent fulfils 88\% of all requests in all scenarios, but its bad action ratio increases from 16\% to 30\% as the network load increases.
The fulfilling ratio of the next-used agent is close to 99\% all the time, but its bad action ratio is between 19\% to 39\%.
In Fig.\ref{fig:rlvalidation}(b) the total rewards for \IP and next-used agents are more than 1.9 times of that of the random agent in the low load scenario.
However, as the network load increases, the total reward for either the random or the next-used agent goes below 0, implying that they have on average a negative impact on the network.
On the other hand, the total reward for the \IP agent always remains positive.

In the \replaced{LoED}{real-world} dataset scenario the \IP and next-used agents perform similarly since the system load is lower than that of our low-load simulated scenario. 
The random agent performs much worse than the other two in terms of the total reward.
Its bad action ratio, however, is similar to the others.
This shows that the random and next-used agents may work well in low load scenarios, but the \IP agent performs better than both regardless of the load on the system.

It is clear from the results above that \AlgoName has more communication opportunities and causes less interference when using \IP over the other policies.
\IP uses a SARSA continuous learning model to train a gateway agent to learn a
model of its local communication traffic. A single model would not work on all gateways due to the multitude of wireless interference and placement issues. \IP also enables a gateway agent to deal with changes in the environment like new nodes, dynamic inter-arrival times or environmental conditions.

\section{Implementation and Evaluation}
\label{sec:evaluation}

In this section we describe our evaluation of \AlgoName. We perform four performance studies;  initially on a small-scale testbed to demonstrate that it works on gateways without making changes to the LoRaWAN protocol. The next two studies are simulations to evaluate how \AlgoName compares against LoRaWAN in highly-dense environments. The final simulation study compares our work against an full-oracle wired centralised solution (WCS).
\\
\textbf{Evaluation Criteria.} We use three evaluation criteria: 
\begin{itemize}[leftmargin=\parindent,topsep=1.5pt]
\item \textbf{Unique messages/node} - The number of unique %
    messages received at the server per node. Each node sends approximately $480$ messages per experiment(depending on start time). Each message is identified by a counter number ${0,\cdots,479}$. A unique message has a unique counter number, and is received only once by the gateway. This is an application level metric and shows how much unique information a server received from a node. 
    \item \textbf{Packet Delivery Ratio (PDR in \%)} - Packet delivery ratio is defined as number of messages acknowledged over the total number of unique messages sent by the node. This metric captures how many messages that needed acknowledgements were successfully acknowledged. 
    \item \textbf{Number of retransmissions (NoReTx)} - The average number of 
    retransmissions needed for messages that require acknowledgements to be acknowledged.
    Reduction in NoReTx implies that nodes have to retransmit less and conserve their energy. 
\end{itemize} 
\textbf{Evaluation algorithms.} We compare \AlgoName with a baseline algorithm LoRaWAN: 
\begin{itemize}[leftmargin=\parindent,topsep=1.5pt]
\item \textbf{LoRaWAN:} This is baseline LoRaWAN that represents a typical use-case for LPWANs. Nodes implement Class A specification\cite{LoRaWAN} and use Adaptive Data Rate(ADR). There is no G2G communication in LoRaWAN. 
\item \textbf{\AlgoName:} This inherits all properties of LoRaWAN and implements the G2G communication.
\end{itemize}

\subsection{Testbed Evaluation}

\subsubsection{\textbf{Setup.}} We evaluated \AlgoName with two overlapping LoRaWAN networks, each with $5$ nodes and $1$ gateway connected to a server. \added[]{The nodes and gateways were deployed in a $150^2$ metre indoor office environment. The nodes were placed about 4 metres apart from each other and the gateways.}
We created collisions by having all of the nodes transmit at the same time, on the same frequency,
using spreading factor $7$, \added[]{and a transmit power of $14$ dBm}. We also reduced the duty-cycle on one of the gateways to $0$\%. The nodes transmitted a message every $20$ seconds and we run the experiment for $5$ minutes and repeat it $10$ times. We report on the average and standard deviation. \\
Our LoRaWAN nodes consisted of an Adafruit Feather M0 RFM95 LoRa node communicating over USB to a host running Linux. All of the hosts were synchronised with NTP, and they instructed the LoRaWAN nodes to transmit at the same time to ensure message collisions. We used $2$ MultiConnect Conduit Gateway as LoRaWAN gateways each with an 868MHz +3dbi whip antenna.
\subsubsection{\textbf{Results.}} 
\begin{figure}[ht!]
    \begin{subfigure}{0.46\textwidth}
      \centering
        \includegraphics[width=\textwidth]{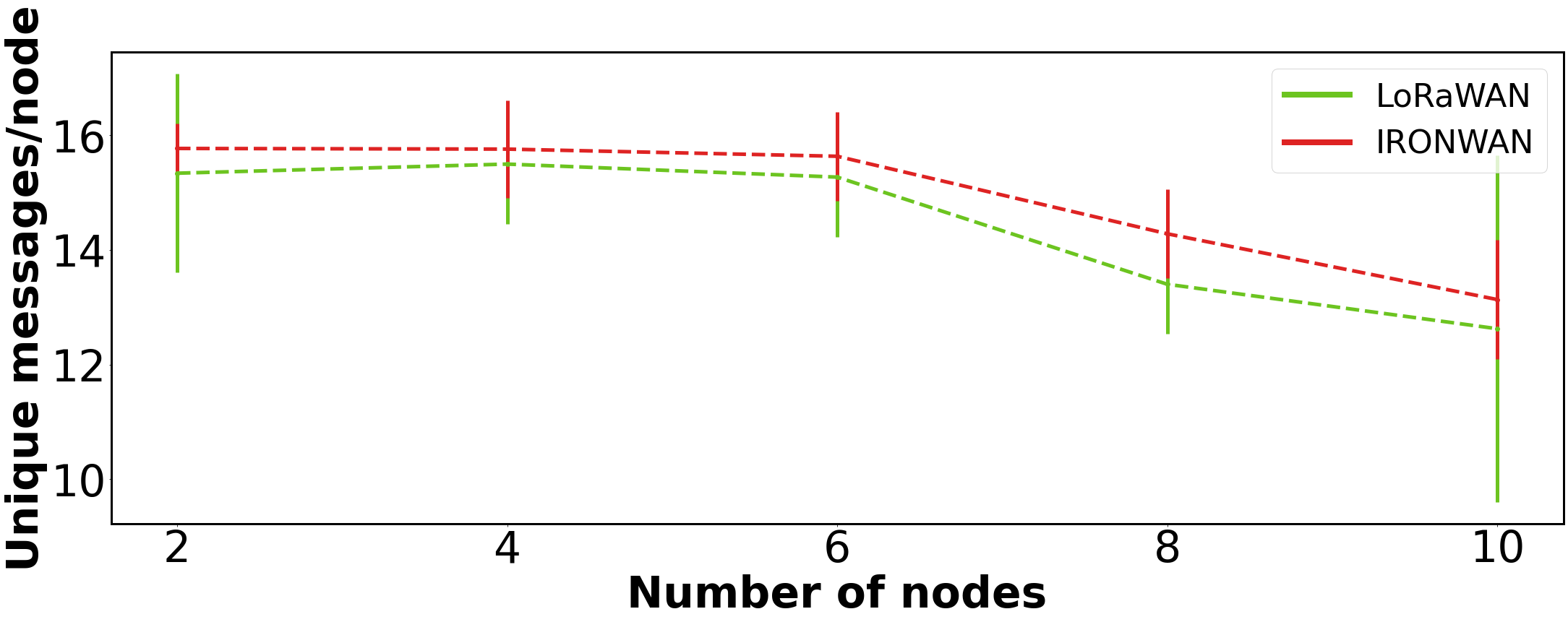}
      \caption{\label{fig:hardware_goodput}}
    \end{subfigure}
    \newline
    \begin{subfigure}{0.46\textwidth}
      \centering
        \includegraphics[width=\textwidth]{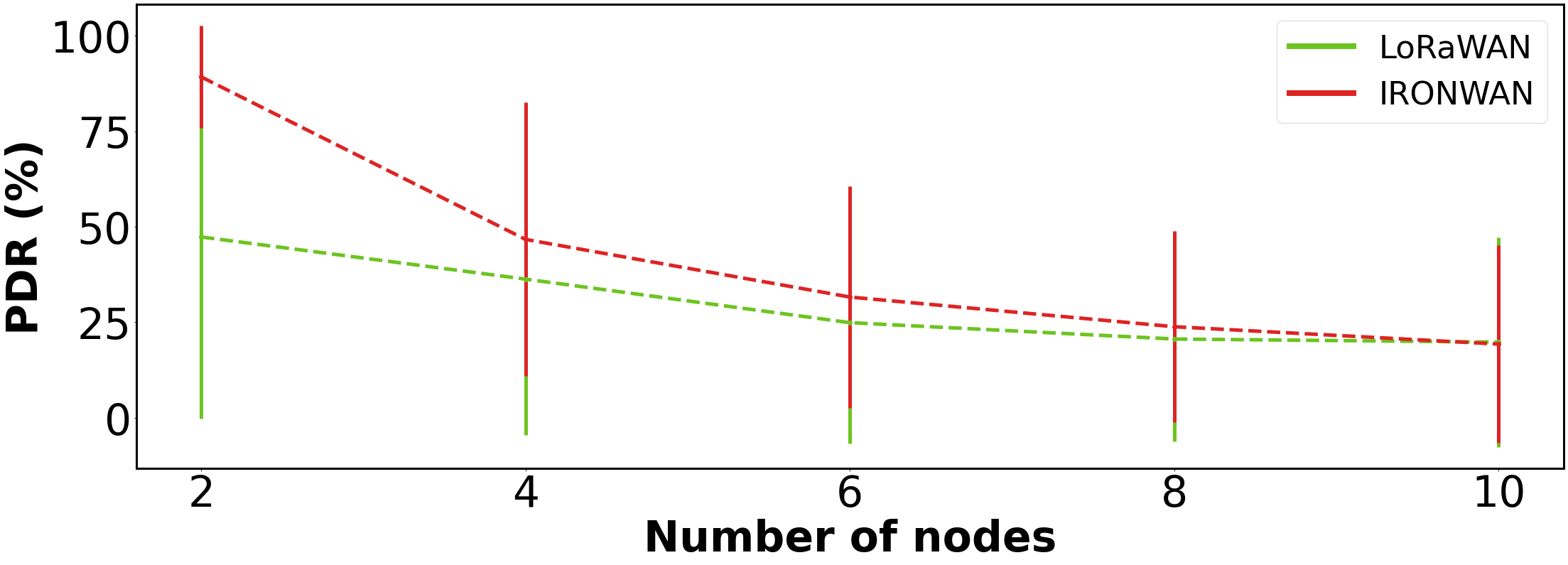}
      \caption{\label{fig:hardware_pdr}}
    \end{subfigure}
    \caption{Testbed evaluation}
    \label{fig:testbedeval}
    \vspace{-5mm}
\end{figure}

Fig.\ref{fig:testbedeval} show results of the testbed experiment. The results show that the average unique messages per node increases by 2-6\% and the average PDR increases by 15-88\%. The results indicate that \AlgoName works on gateways and can improve the total messages received and PDR.

\subsection{Large-scale Simulation Evaluation}

Next, we perform large-scale simulations to study how \AlgoName works in dense urban scenarios. 

\textbf{Setup.} We evaluate \AlgoName on the FLORA\cite{8406255} LoRaWAN
simulator using OMNeT++ and the INET framework. We simulate $1000$ nodes
uniformly deployed in a $4km^2$ simulation area.
We base the density of the gateway deployment on an experimental study where an average of $2-3$ gateways should receive a message for any node with a gateway density of one every $1.5km^2$ \cite{lorawancapacity}. We simulated the use of six,
eight, and ten gateways, $G \in {6,8,10}$, to give an average of $1.5, 2, 2.5$
gateways per network.  Our simulated networks had $G$ gateways connected to $4$
independent networks with their own servers.  The nodes transmit a new message
every three minutes (maximum of $480$ messages). The experiment is run for $24$
hours of simulated time. We define the load on the network as the percentage of
nodes that require acknowledgements for all of their messages. We evaluate
three loads on the network: low, medium and high that correspond to 10,50 and 90\% of nodes requiring acknowledgements.

\subsubsection{\textbf{Study 1: Impact of increasing gateways}}
In the first study, we compare LoRaWAN and \AlgoName. 
We increase the number of gateways and the load on the network. 

\begin{figure}[ht!]
\begin{subfigure}{0.48\textwidth}
  \centering
    \includegraphics[width=\textwidth]{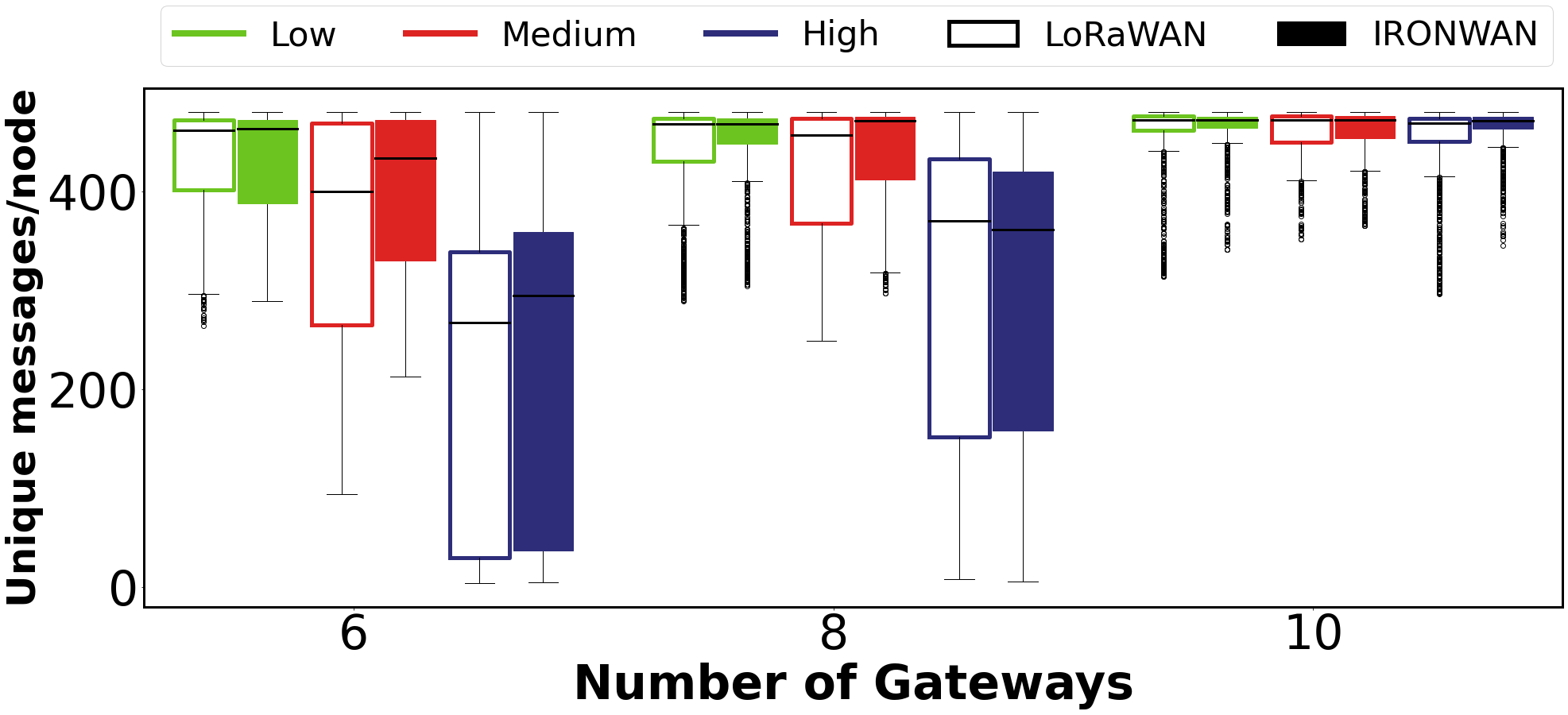}
  \caption{\label{fig:goodput_study_2}}
\end{subfigure}
\newline
\begin{subfigure}{0.48\textwidth}
  \centering
    \includegraphics[width=\textwidth]{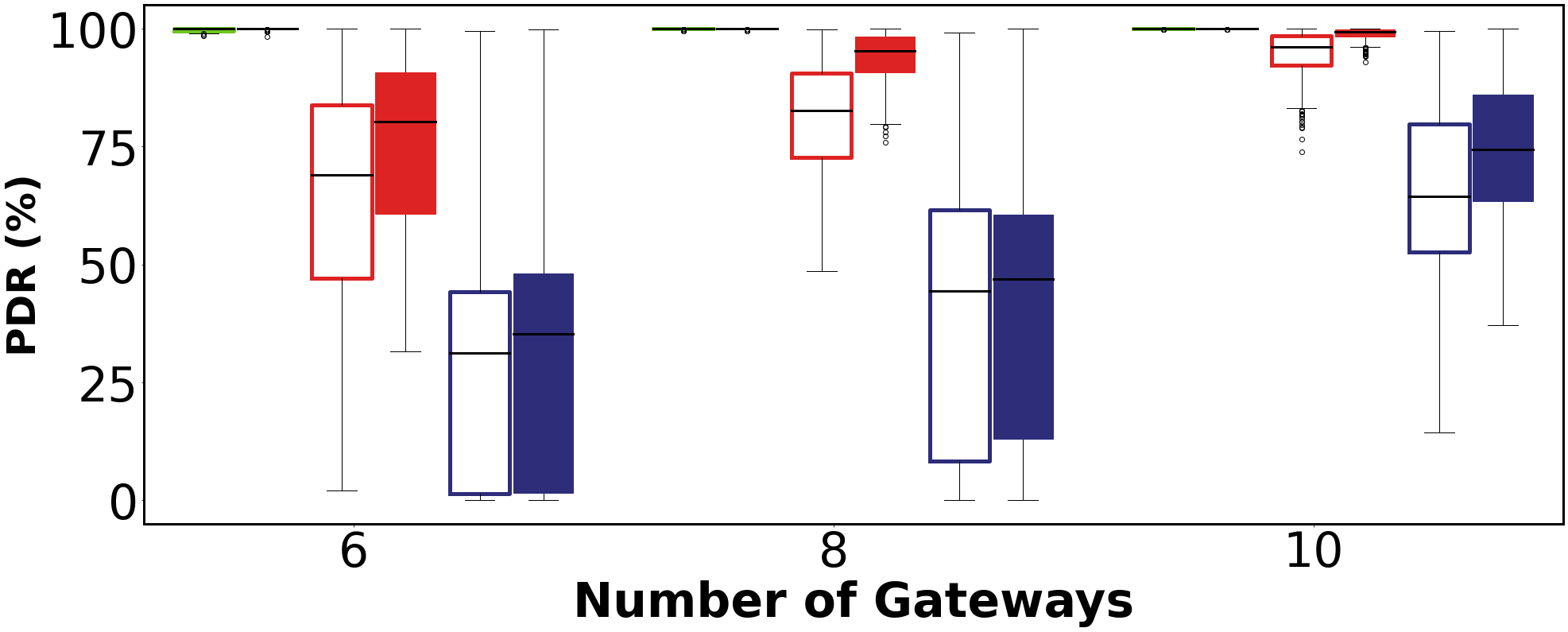}
  \caption{\label{fig:pdr_study_2}}
\end{subfigure}
\caption{Evaluation of LoRaWAN and \AlgoName with increasing load and number of gateways}
\vspace{-3mm}
\end{figure}

\textbf{Unique Messages Received and PDR.} Fig.\ref{fig:goodput_study_2} shows that the
total messages received increases with an increase in gateways for both LoRaWAN and \AlgoName.
We see \AlgoName receives 12\% more messages than LoRaWAN for a
medium load scenario with 6 gateways. For other scenarios, the average
total messages received by \AlgoName is from 1-7\% better than that for LoRaWAN. 
We attribute the low throughput gain
difference to the LoRaWAN Adaptive Data Rate which reduces the transmission
power to reduce the number of nodes heard at multiple gateways.  
\AlgoName increases the minimum number of messages received from a node in all but the
high load with 8 gateways and reduces the 25th percentile (lower line of the
box) for all scenarios. 
\AlgoName reduces the starvation of nodes and the servers have more information from every node. 

Fig.\ref{fig:pdr_study_2} shows that the average PDR in low-load scenarios is
above 99.7\% and 99.9\% for LoRaWAN and \AlgoName respectively. With a medium-load
\AlgoName has 11\%,12\% and 5\% higher PDR for 6,8 and 10 gateways. A similar but smaller PDR performance
increase is seen for \AlgoName when the network has a high load. The minimum
PDR does not change in low-load scenarios. For medium load \AlgoName has a
minimum PDR that is better than the LoRaWAN's minimum PDR by 28\% for 6
gateways, 27\% for 8 gateways and 20\% for 10 gateways. \AlgoName's minimum PDR for a high network load with $10$ gateways is 23\% higher than that of LoRaWAN. The results show that networks with a medium or high load have a higher PDR with \AlgoName than with LoRaWAN. We see that for networks with a high load and 10 gateways \AlgoName enables gateways to handle more
acknowledgements and improve its minimum and average PDR.

Fig.\ref{fig:goodput_study_2} and Fig.\ref{fig:pdr_study_2} show that as the
load on the network increases, \AlgoName increases the reliability by handling
more acknowledgements which reduces the load on the network. We then see the impact of \AlgoName on NoReTx. 

\begin{figure}[ht!]
    \centering         
    \includegraphics[width=0.47\textwidth]{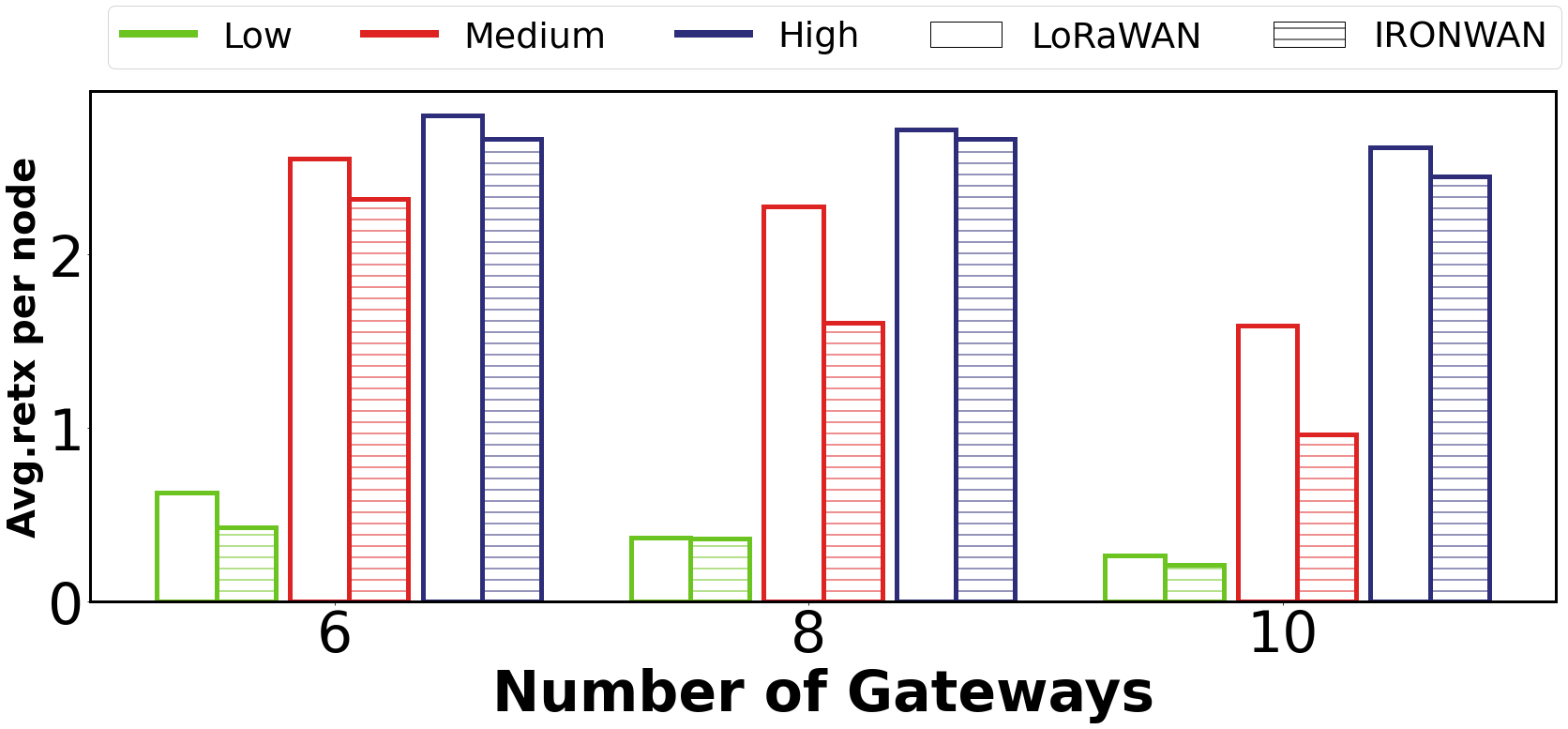}
    \caption{Average number of retransmissions}
    \label{fig:retransmissions}
    \vspace{-3mm}
\end{figure}

\textbf{Number of retransmissions} \AlgoName and LoRaWAN have very similar
messages received per node as seen in Fig.\ref{fig:goodput_study_2}. The
difference is in the number of retransmissions required to achieve the messages received. Fig.\ref{fig:retransmissions} shows the NoReTx per node. 
\AlgoName reduces the NoReTx in low load scenarios by 1\%,21\% and 32\%, for medium load by 8\%,29\%,39\%, and for high load by
2\%,5\% and 6\% when compared to LoRaWAN. This shows that \AlgoName can
achieve similar performance to LoRaWAN while consuming lesser energy in low and medium load scenarios.  

\begin{figure}[ht!]
    \centering         
    \includegraphics[width=0.47\textwidth]{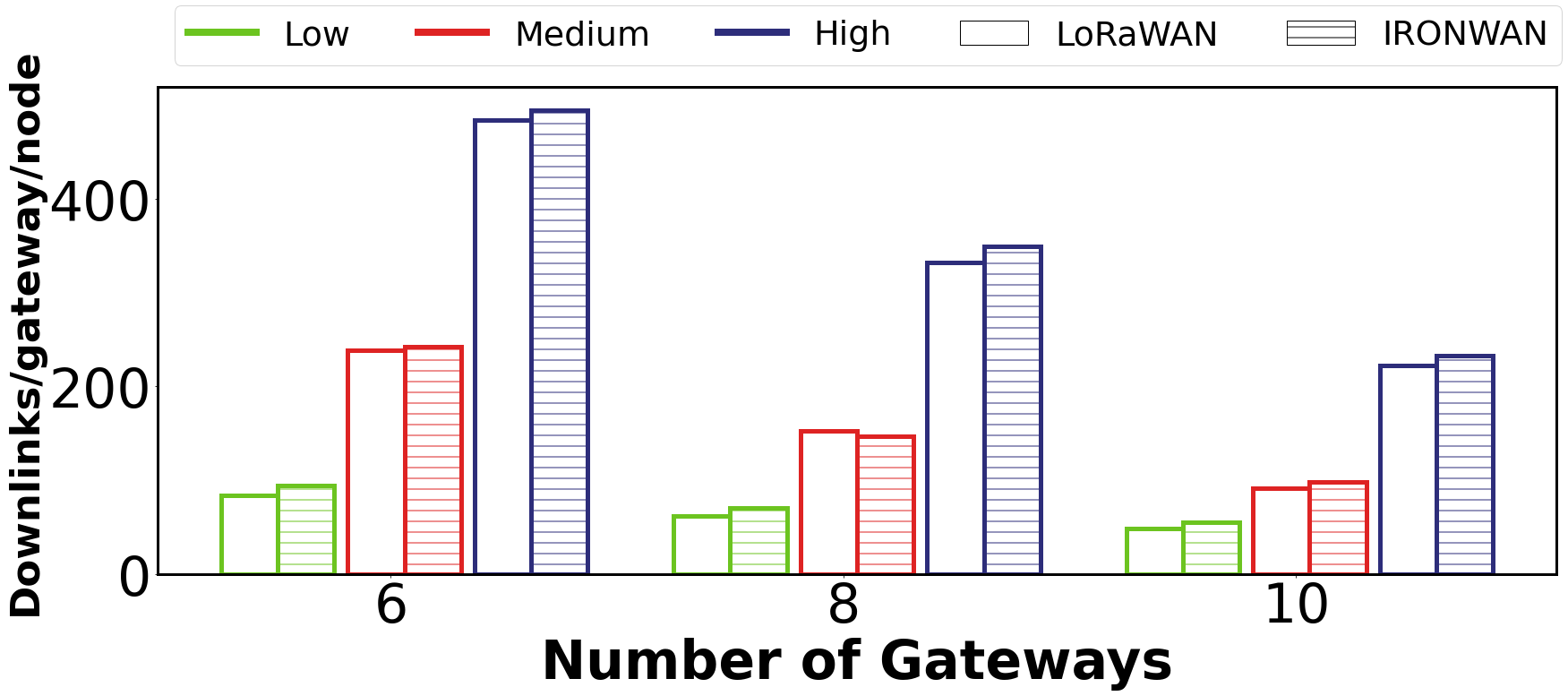}
    \caption{Overhead for \AlgoName compared to LoRaWAN}
    \label{fig:overhead}
    \vspace{-5mm}
\end{figure}

\textbf{Overhead of \AlgoName} In Fig.\ref{fig:overhead} we see the
additional communication overhead caused by the G2G communication used in \AlgoName when compared to LoRaWAN. \AlgoName's overhead is the number of messages transmitted on bands 0 and 1 and LoRaWAN's overhead is only messages transmitted on band 1. \AlgoName's overhead is 12-14\% for low load scenarios, 0-7\% in medium load, and 2-5\% in
high load scenarios. An interesting observation is that when \AlgoName is used
in a network with a medium load and 8 gateways it has its highest gain in PDR
(Fig.\ref{fig:pdr_study_2}) and transmits less messages per node. This
occurs because \AlgoName provides a better redistribution of resources and
reduces acknowledgements. The results show that overhead of \AlgoName is not
high and that it uses spare gateway duty-cycles to the benefit of all
networks.

\subsubsection{\textbf{Study 2: Impact of increasing number of retransmissions}}

In the second study, we test the effect of varying the maximum number of
retransmissions on LoRaWAN and \AlgoName. LoRaWAN allows users to choose a
policy that limits the maximum number of
retransmissions($retx$). LoRaWAN specifications recommend a retransmission
limit of $8$ which is what we use in all other experiments. It was shown in
\cite{8407095} that increasing the number of retransmissions increases the
probability of lost packets. With this experiment, we study the effects of
varying the retransmission limit on the performance of \AlgoName.

\begin{figure}[ht!]
    \begin{subfigure}{0.48\textwidth}
    \includegraphics[width=\textwidth]{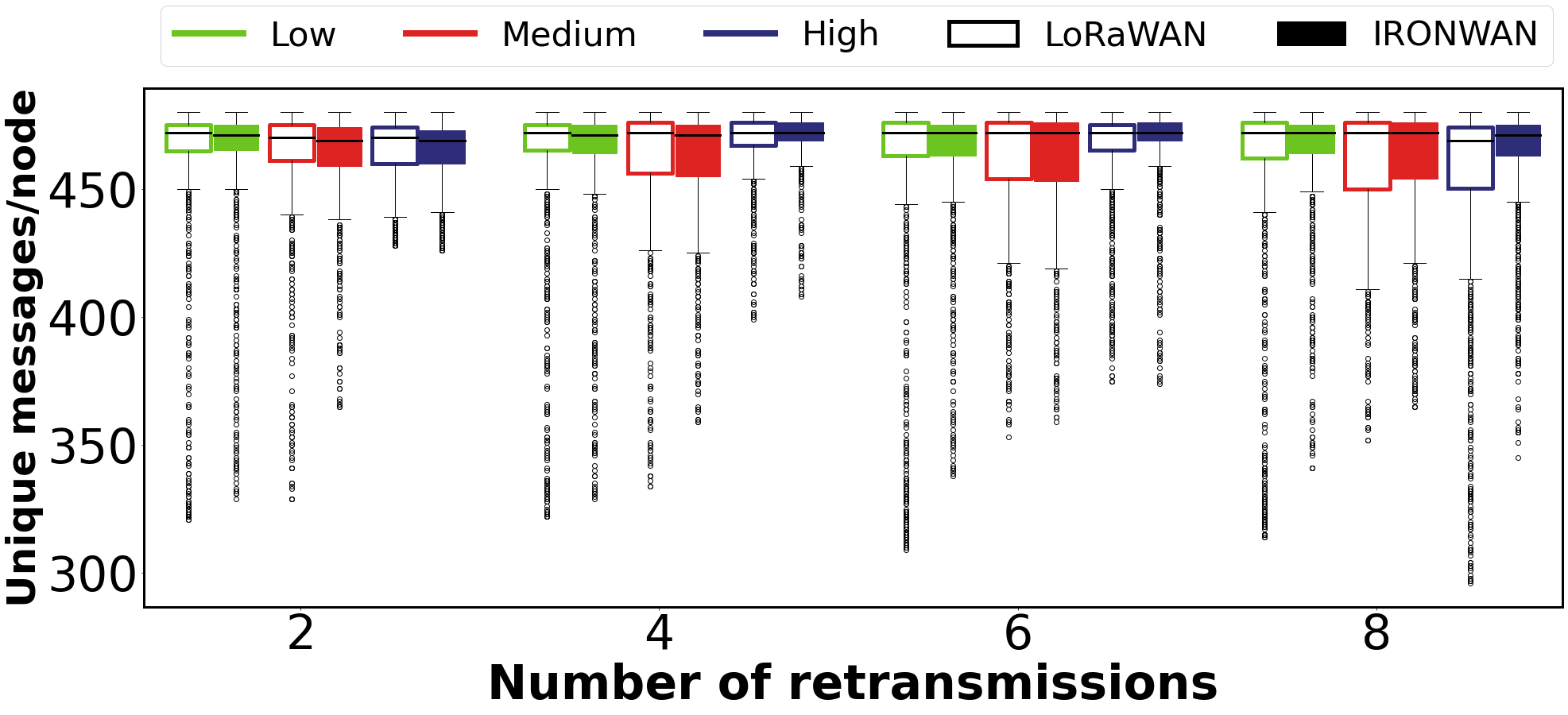}
    \caption{\label{fig:goodput_study_3}}
    \end{subfigure}
    \begin{subfigure}{0.48\textwidth}
    \includegraphics[width=\textwidth]{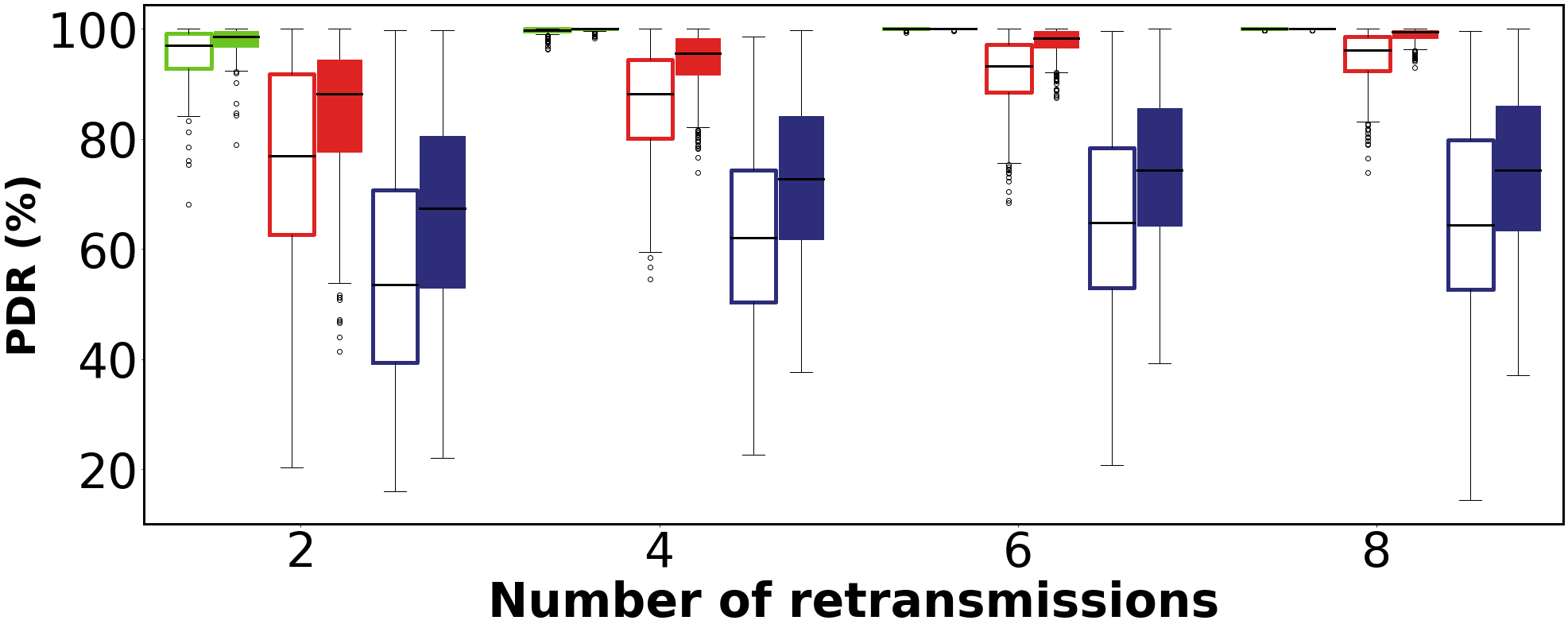}
    \caption{\label{fig:pdr_study_3}}
    \end{subfigure}
    \caption{Varying maximum number of retransmissions}
    \label{fig:retransmission_limit}
    \vspace{-5mm}
\end{figure}

\underline{Results.} Fig.\ref{fig:retransmission_limit} shows the unique
messages received and PDR for a maximum of 2,4,6 and 8 retransmissions in a network with $10$ gateways distributed between $4$-servers. 
The total messages received is similar for all networks loads. This allows us
to see how the PDR changes to achieve the same performance.
A clear trend emerges where the PDR reduces to 78\% for a medium load and 50\%
for a high load for $2$ retransmissions for LoRaWAN. For \AlgoName under the
same conditions the PDR only drops to 90\% and 63\%.
Another observation is that \AlgoName
with $retx$ retransmissions has a 5\% to 10\% higher PDR than
LoRaWAN with $retx + 2$ retransmissions. Instead of increasing $retx$,
\AlgoName could be used instead which would reduce the load on the network and
the number of messages sent by the nodes 
by replacing node retransmissions with G2G messages requests.
This can also be seen in Fig.\ref{fig:pdr_study_3}. In all
scenarios, the average PDR of the system increases by up to 20\% and the
minimum PDR increases in range of 25-160\% for medium and high load scenarios.
Increasing the retransmissions increases the load on the network and does not
significantly increase the PDR. This is evident from the high-load scenario
where the average PDR is always higher in \AlgoName compared to cases in which
the retransmission limit is increased. This study shows that \AlgoName
increases the total messages received and PDR of the system when compared to LoRaWAN with an increasing
retransmission limit.

\subsubsection{\textbf{Study 3: Comparison to state-of-art}}
In our final study, we compare \AlgoName with two state-of-art systems. As both
of the algorithms under comparison have components that tackle
different issues, we focus only on the parts of solution that are comparable to
\AlgoName. The two solutions are: 

\textbf{1. Wired centralised server (WCS).} WCS is a centralised system that facilitates message exchange between multiple
servers. In WCS every time a server
receives a message, it forwards messages not destined for it to all other
servers. So, the servers have information from all gateways that received their message, and they can also use those gateways for acknowledgements.
However, they only choose to use those gateways if none of their gateways
had received a message. We calculate the overhead of WCS as the
number of times it used a message received from another server or used another
server for acknowledgements. This is the most significant difference between a network with 1-stakeholder and WCS. The closest thing to WCS is PacketBroker, however, there are no documentations or publications that describe how they work. 

\begin{figure}[ht!]
\begin{subfigure}{0.48\textwidth}
  \centering
    \includegraphics[width=\textwidth]{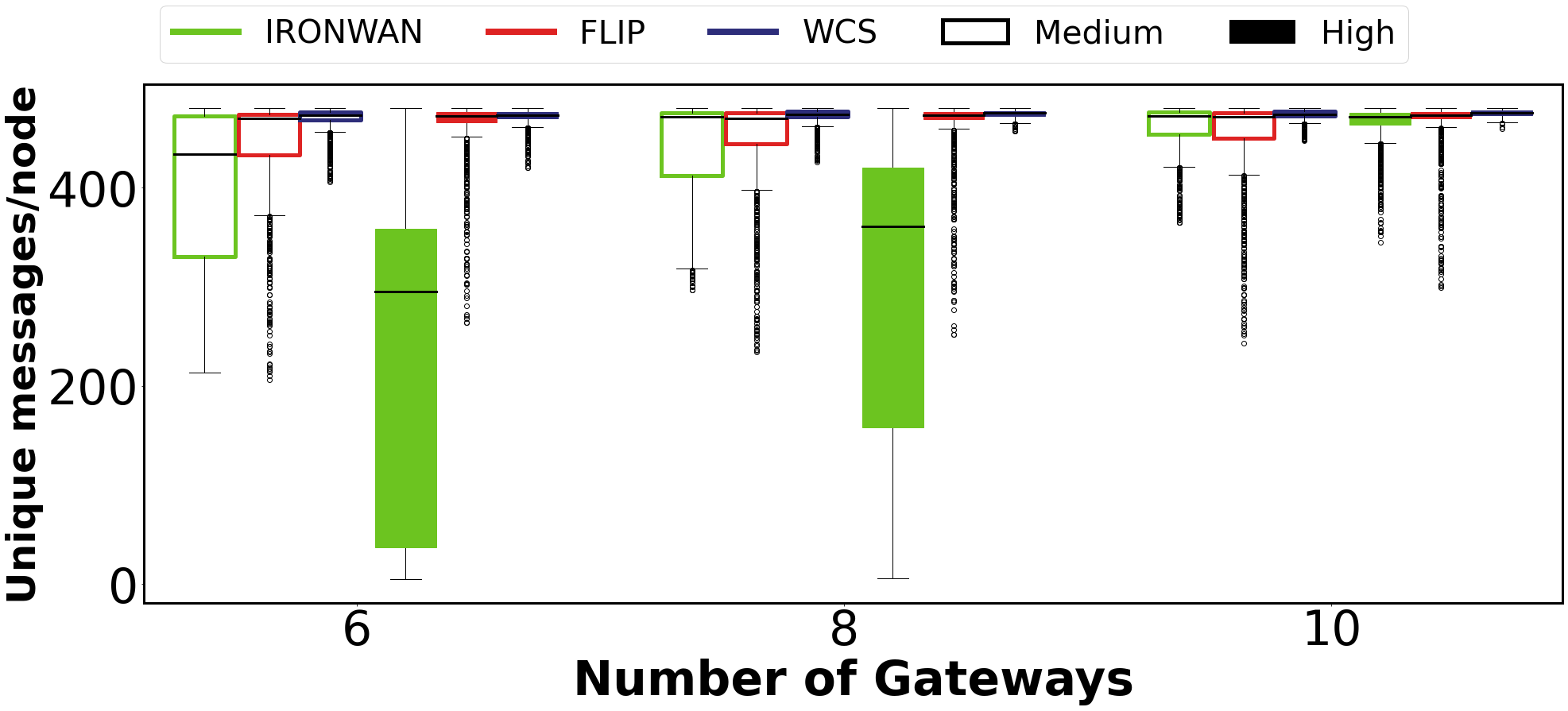}
    \caption{\label{fig:goodput_study_4}}
\end{subfigure}
\newline
\begin{subfigure}{0.48\textwidth}
  \centering
    \includegraphics[width=\textwidth]{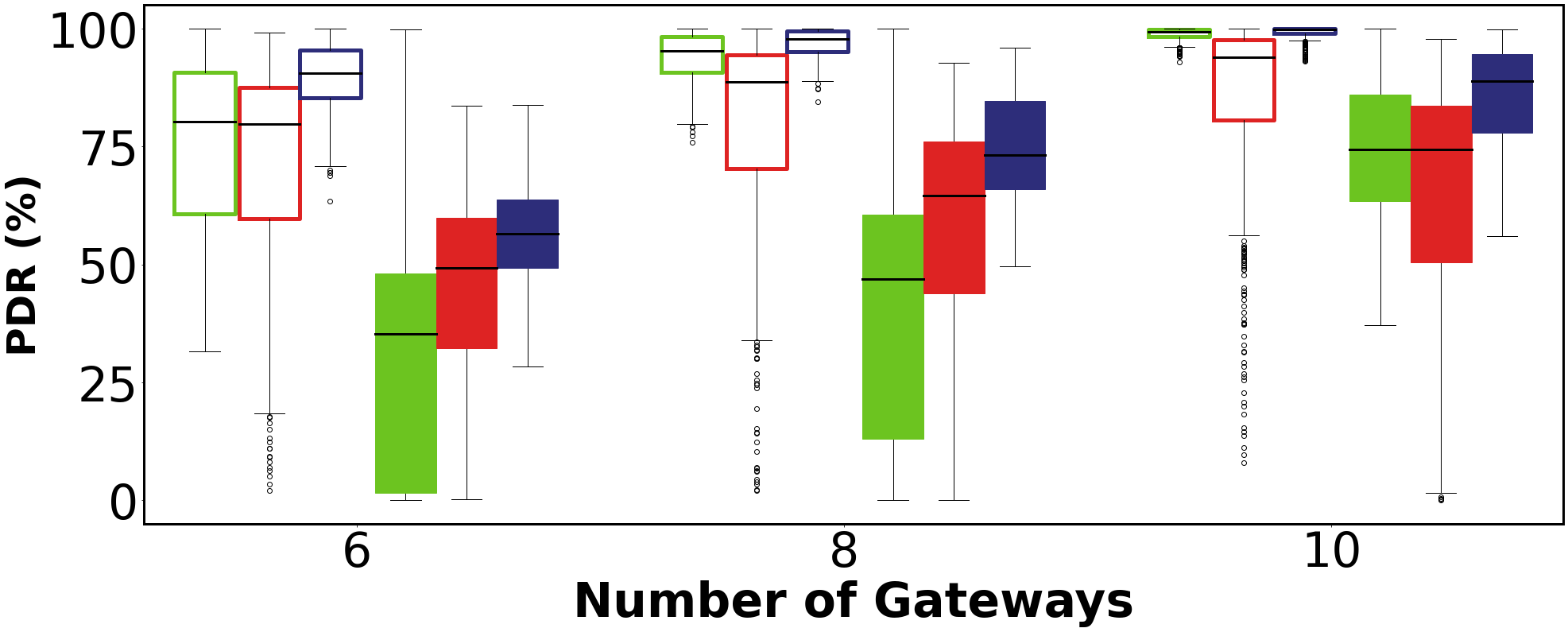}
  \caption{\label{fig:pdr_study_4}}
\end{subfigure}
\begin{subfigure}{0.48\textwidth}
  \centering
    \includegraphics[width=\textwidth]{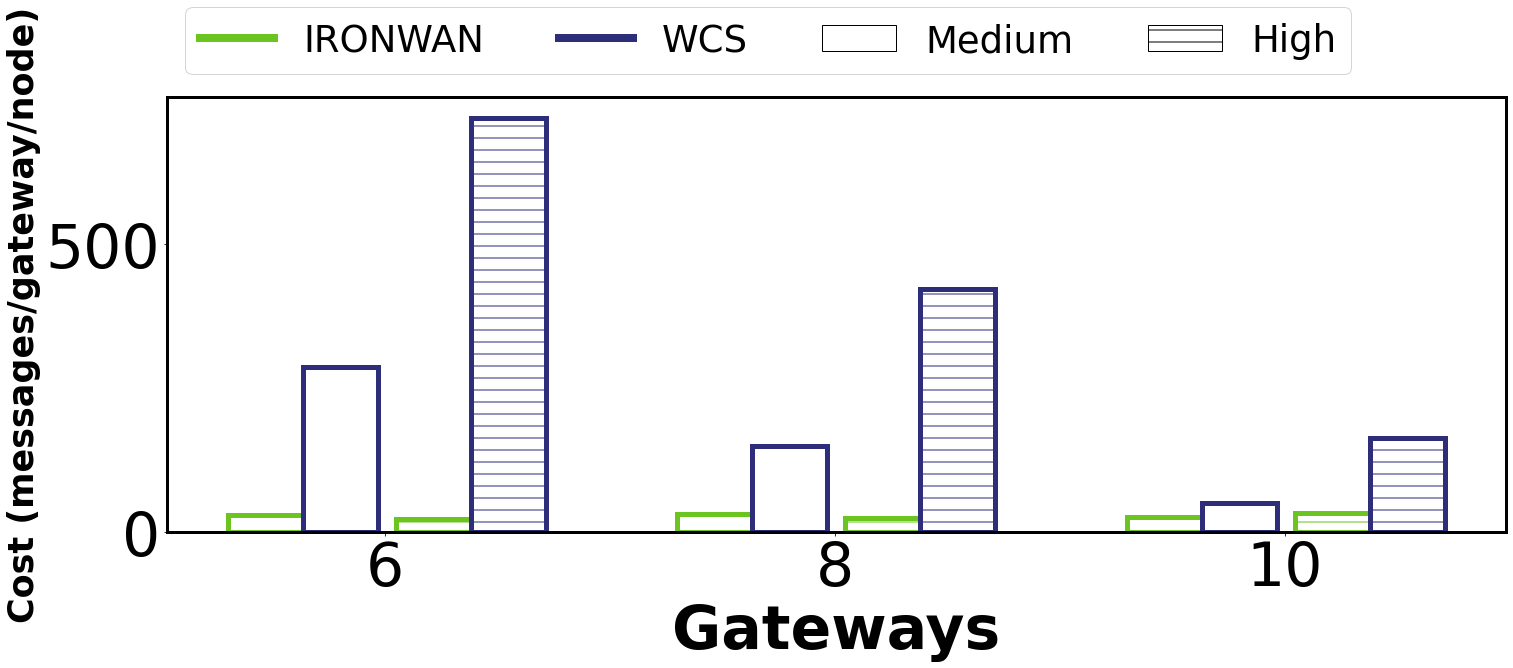}
  \caption{\label{fig:overhead_study_4}}
\end{subfigure}
\caption{Comparison of \AlgoName, FLIP and WCS}
\vspace{-5mm}
\end{figure}

\textbf{2. FLIP.} FLIP is a peer-to-peer network between
gateways to distribute the load of the network. FLIP does this by
minimising Shannon's entropy and coming to a consensus about what nodes are
handled by which gateways. For our comparison, we implement a centralised
server that receives all of the messages and assigns nodes to gateways such
that the entropy is minimised. We use this technique as it mimics the
operation of FLIP which minimises entropy by sharing the load. \\
\underline{Results.} Fig.\ref{fig:goodput_study_4} and
Fig.\ref{fig:pdr_study_4} show the results of our comparison. We only show results
for medium and high load as all three algorithms perform similarly under the
low-load scenario. WCS always has the highest number of unique messages received and PDR as it provides the behaviour of an unpartitioned network. The cost required by
WCS to achieve this can be seen in Fig.\ref{fig:overhead_study_4}.
For a medium load scenario, the total messages received by \AlgoName is as good
as FLIP with 6 gateways and outperforms it when 8 and 10 gateways are used. The
total messages received by FLIP is higher than that for \AlgoName in the 6 and
8 gateway scenarios.  \AlgoName receives more messages than FLIP in the 10
gateway scenario. The PDR for medium load follows a similar trend, where
\AlgoName has a higher PDR and much higher minimum PDR when compared to FLIP
and is comparable to WCS in the 10 gateway scenario. For a high load
scenario, \AlgoName has a lower PDR when compared to FLIP and WCS,
however, it increases the maximum PDR from 80\% in FLIP and
WCS to 100\%. \AlgoName's PDR is higher than FLIP in 10 gateway
scenario and closer to WCS's. Fig.\ref{fig:overhead_study_4} shows that the cost of \AlgoName is significantly lower than WCS in all scenarios where cost for \AlgoName is number of messages it requested or responded for other networks. 

The results of our evaluation show that \AlgoName increases total messages
received by a limited amount while increasing the PDR in most scenarios. It also increases the minimum PDR in all scenarios. \AlgoName also reduces the number of
retransmissions in all scenarios with minimal overhead per gateway. \AlgoName
does not perform as well as WCS (which is similar to an unpartitioned network) but
performs better than FLIP in terms of PDR. Finally, our results show that \AlgoName is a candidate solution for dealing with multi-owner overlapping networks commonly found in LoRaWAN.

\section{Limitations and Future Work}
\label{sec:limitations}

\AlgoName currently has no idea about how many gateways are in its communication range. A neighbourhood discovery protocol could help gateways form a map of its neighbouring gateways. This map could then be used to create trust and incentivisation schemes to deal with free-loading or malicious gateways. 
The G2G messages are not currently encrypted nor authenticated, we
leave that for further work. We plan to use neighbour disovery algorithms to create keys for G2G authentication in a way that would not require network owners to coordinate thus retaining the deployment security and simplicity of \AlgoName. 
\\
\Arr has been designed for periodic data, it could be extended to deal with event-based data which would allow our system to be used in more scenarios.
We have tested \IP for a limited time and over-time the environment it operates in can change. 
A real-time parameter tuning and periodic recalibration needs to happen to deal with drastic changes in the wireless environment.

\section{Related Work}
\label{sec:related}
In this section we review current research that tackles aspects of the multi-gateway and multi-owner networks problem in LoRaWAN.

\textbf{Benefits of multiple gateways:} Authors in \cite{8761157} showed that increasing LoRaWAN gateways from one to four improved the messages that were acknowledged from 24\% to 40\%. The authors of \added[]{\cite{dongare2018charm,10.1145/3356250.3360024,Liu2020}}
used multiple gateways to directly forward all of signals received from all of the nodes in their range to a central server where they are combined and decoded. This is an orthogonal approach to ours as
it changes the role that gateways play in a LoRaWAN network by offloading decoding of packets to a server which requires precise time-synchronisation, high bandwidth wired links and most importantly all gateways talking to the same server. 
To the best of our knowledge,
all previous work has assumed that multi-gateway LoRaWAN networks share the
same server which is contradictory to LoRaWAN deployment methods where multiple overlapping networks exist. Our work removes this assumption and allows networks to retain
their autonomy while still working together to improve performance for all.

\textbf{Downlink traffic:} The authors of \cite{8407095,alex2017does} examine the effects of downlink traffic in LoRaWAN and demonstrate its inability to handle high amount of
downlink traffic. A number of solutions have been proposed
\cite{8761157,bhatia2020control,oh2018trilo} to increase the reliability of
LoRaWAN by balancing the load, scheduling traffic or controlling access
to the channel using queuing systems. All of this assumes a single network
using a \textit{single server}. \AlgoName addresses the more realistic
scenario of multiple overlapping networks and solves these by enabling communication between the gateways of different networks.

\textbf{Merging overlapping networks:} The most notable work that tackles the problem of overlapping networks is FLIP\cite{FLIP}. FLIP federates gateways of multi-stakeholder overlapping networks. The federated gateways
distribute the nodes in overlapping communication regions amongst each other at initialisation. The gateways then handle the communication with the nodes assigned to them as they also hold the encryption keys for that node. The gateways then forward the received messages to the gateway that owns the node which in-turn forwards it to the server. The federated gateways share decryption keys for the
allocated nodes which poses a threat in the case of malicious gateways. FLIP assumes a wired internet connection for G2G communication which may not be available, reliable, or secure. In \AlgoName gateways do not require the keys of neighbour nodes. \AlgoName uses the wireless spectrum
overcoming the problem of an unreliable wired
connection\cite{unreliableinternet} or unavailable backhaul networks in hazardous scenarios. As shown in Sec.\ref{sec:evaluation}, \AlgoName performs better than FLIP in most scenarios without having any security issues.

\section{Conclusion}
\label{sec:conclusion}

In this paper, we present \AlgoName and its novel components \Arr and \IP. \AlgoName leverages overlapping LoRaWAN networks to enable wireless gateway-to-gateway communication, reducing the negative effects of node-to-gateway message collisions and efficiently sharing gateway-to-node communication capacity. Both \Arr and \IP are new approaches to solve these problems. We evaluate the effectiveness of \AlgoName in simulation and a testbed experiment. Our results show that \AlgoName outperforms LoRaWAN in low-load and medium-load scenarios (can be considered typical use-cases of LoRaWAN) and achieves comparable performance in the high-load scenario. 
\AlgoName also improves the messages received per node and the packet delivery ratio while reducing the number of retransmissions required and enables gateways to acknowledge messages when they have exhausted their communication duty-cycle. 
Ultimately, \AlgoName is a suitable candidate to leverage overlapping networks to increase the reliability for all participating networks in LoRaWAN deployments.

\bibliographystyle{IEEEtran}
\bibliography{main}

\end{document}